# SPECTRAL EVOLUTION OF AN EARTH-LIKE PLANET

LISA KALTENEGGER

Harvard Smithsonian Center for Astrophysics, 60 Garden Street, 02138 MA Cambridge, USA; lkaltenegger@cfa.harvard.edu

WESLEY A. TRAUB

Jet Propulsion Laboratory, M/S 301-451, 4800 Oak Grove Drive, Pasadena, CA 91109, USA; also Harvard Smithsonian Center for Astrophysics, 60 Garden Street, 02138 MA Cambridge, USA

AND

KENNETH W. JUCKS

Harvard Smithsonian Center for Astrophysics, 60 Garden Street, 02138 MA Cambridge, USA

## ABSTRACT

We have developed a characterization of the geological evolution of the Earth's atmosphere and surface in order to model the observable spectra of an Earth-like planet through its geological history. These calculations are designed to guide the interpretation of an observed spectrum of such a planet by future instruments that will characterize exoplanets. Our models focus on planetary environmental characteristics whose resultant spectral features can be used to imply habitability or the presence of life. These features are generated by $H_2O$, $CO_2$, $CH_4$, $O_2$, $O_3$, $N_2O$, and vegetation-like surface albedos. We chose six geological epochs to characterize. These epochs exhibit a wide range in abundance for these molecules, ranging from a $CO_2$ rich early atmosphere, to a $CO_2/CH_4$-rich atmosphere around 2 billion years ago to a present-day atmosphere. We analyzed the spectra to quantify the strength of each important spectral feature in both the visible and thermal infrared spectral regions, and the resolutions required to optimally detect the features for each epoch. We find a wide range of spectral resolutions required for observing the different features. For example, $H_2O$ and $O_3$ can be observed with relatively low resolution, while $O_2$ and $N_2O$ require higher resolution. We also find that the inclusion of clouds in our models significantly affects both the strengths of all spectral features and the resolutions required to observe all these.

*Subject headings*: extrasolar planets, spectra, biomarkers, exoplanets

## 1. INTRODUCTION

Over 200 exoplanets already have been detected, and hundreds, perhaps thousands more, are anticipated in the coming years. The detection and characterization of these exoplanets will begin to fill in a conspicuous gap in the astrophysical description of the universe, specifically the chain of events occurring between the first stages of star formation and the evident, now abundant, mature planetary systems. The nature of these planets, including their orbits, masses, sizes, constituents, and likelihood that life could develop on them, can be probed by a combination of observations and modeling.

The present stage of exoplanet observations might be characterized as one in which information is being gathered principally by ***indirect*** means, whereby the photons that we measure are from the star itself, or a background star, or a mixture of the star and planet. Indirect techniques include radial velocity, micro-lensing, transits, and astrometry. These indirect observations are of great value, giving us measures of the planet mass, orbital elements, and (for transits) the sizes as well as indications of the constituents of the extreme upper atmospheres. Of special interest is the detection of sodium in the upper atmosphere of HD209458b (Charbonneau et al. 2002), and because of its implication that Earth-mass planets might be common, we note the detection of a 5.5 Earth-mass planet by microlensing (Beaulieu et al. 2006).

In the next stage of exoplanet observations, we may hope to have ***direct*** observations, in which most of the measured photons are reflected or emitted by the planet itself. Direct techniques include coronagraphic imaging at visible wavelengths, and interferometric imaging in the thermal infrared. With direct photons from the visible and thermal infrared, and depending on the particular cases, we can characterize a planet in terms of its size, albedo, and, as will be discussed in this paper, its atmospheric gas constituents, total atmospheric column density, clouds, surface properties, land and ocean areas, general habitability, and the possible presence of signs of life. At higher signal-to-noise ratios we will also be able to measure rotation period, weather variability, the presence of land-plants, and seasons.

As discussed in Traub et al. (2006), full characterization requires the synergy of having both direct and indirect measurements. Note that direct detection of photons from giant exoplanets can be implemented using current space based telescopes like Hubble and Spitzer. Such studies have led to the detection of infrared emission from two transiting hot Jupiters (Deming et al. 2005; Charbonneau et al. 2005) where the planetary signal is the difference between the flux from a star plus planet versus the flux from the star alone. Photons from Earth-like planets in the habitable zone around their parent star are beyond the capabilities of these telescopes and require future missions.

We anticipate that future direct observations will be carried out from space. The thermal infrared concepts, the Terrestrial Planet Finder Interferometer (TPF-I) and Darwin missions, and the visible wavelength concepts, the Terrestrial Planet Finder Coronagraph (TPF-C), are designed to detect terrestrial exoplanets, and to measure the color and spectra of terrestrial planets, giant planets, and zodiacal dust disks around nearby stars (see, e.g., Beichman et al. 1999; 2006; Fridlund 2000; Kaltenegger 2004; Kaltenegger & Fridlund 2005; Borde & Traub 2006). These missions have the explicit purpose of detecting other Earth-like worlds, analyzing their characteristics, determining the composition of their atmospheres, investigating their capability to sustain life as we know it, and searching for signs of life. These missions also have the capacity to investigate the physical properties and composition of a broader diversity of planets, to understand the formation of planets and interpret potential biosignature compounds.

The range of characteristics of planets is likely to exceed our experience with the planets and satellites in our own Solar System. Earth-like planets orbiting stars of different spectral type might evolve differently (Selsis 2000; Segura et al. 2003, 2005). A crucial factor in interpreting planetary spectra is the point in the evolution of the atmosphere when its biomarkers become detectable.

Studies of individual constituent effects on the spectrum and resolution estimates were previously discussed for an Earth-like planet in Des Marais et al. (2002). These calculations were for a current atmospheric temperature structure, but different abundances of chemical species. The work presented here is complementary to the results discussed in Des Marais et al. (2002) because it explores the evolution of the expected spectra of the Earth and its biomarkers over geological timescales. It discusses the signatures found and the resolution needed to detect spectral features and biomarkers throughout Earth's evolution until the present day.

Spectra of the Earth exploring temperature sensitivity (a hot house and cold scenario) and different singled out stages of its evolution have been published previously (e.g., Schindler and Kasting, 2000; Selsis, 2000; Pavlov 2000; Traub & Jucks 2002; Segura et al. 2003; Kaltenegger 2006; Meadows, 2006). Some of those studies focus on single stages in Earth evolution with a single surface type, and/or no clouds, however Traub and Jucks (2002), Kaltenegger (2006) and Meadows (2006) include cloud models and varying surface materials.

The work presented here is a more comprehensive study of the evolution of the Earth's atmosphere over geological time. It improves on these preliminary studies by exploring more epochs, treating the surface more realistically, including the effect of clouds on the spectra, and quantifying the spectral resolution required to detect habitability and biosignatures as a function of time.

In this paper we concentrate on the example of Earth by establishing models for its atmosphere and detectable biomarkers over its evolution history. The work reported here is focused on providing future missions with information to aid their design, and to interpret their results. From the spectra one can screen exoplanets for habitability. The Earth-Sun intensity ratio is about $10^{-7}$ in the thermal infrared (~10 μm), and about $10^{-10}$ in the visible (~0.5 μm). Both spectral regions contain atmospheric bio-indicators: $CO_2$, $H_2O$, $O_3$, $CH_4$, and $N_2O$ in the thermal infrared, and $H_2O$, $O_3$, $O_2$, $CH_4$ and $CO_2$ in the visible to near-infrared. The presence or absence of these spectral features will indicate similarities or differences with the atmospheres of terrestrial planets. We adopt spectral ranges of 5-20 μm in the thermal infrared, and 0.5-1.1 μm in the visible, as proposed for future mission concepts to detect biomarkers (see, e.g., Des Marais et al. 2002; Fridlund 2000).

The organization of this paper is as follows. Section 2 outlines our spectroscopic and atmospheric models, and gives the essential elements of our current understanding of the major epochs of geologic development of the Earth, as they relate to surface and atmospheric properties governing the visible and thermal infrared spectra of the integrated planet. It presents our estimates of the corresponding atmospheric gas and temperature profiles, and also relevant cloud and surface conditions. It gives three examples of Earth-observation validations of our models, for the visible, near-infrared, and thermal infrared. Section 3 combines the preceding models of the six Earth-epochs with our spectroscopic and atmospheric models, to produce visible and thermal infrared spectra for each epoch. The effects of time variations of abundances, clouds, and spectral resolution are presented. The equivalent width and the resolution needed to detect the different features for future space missions are given and their importance discussed. The spectral evidence for habitability and signs of life are presented. Section 4 discusses the results of our models and the spectral sensitivity to clouds, surfaces and planets. Section 5 summarizes our major points, and suggests future work.

## 2. SPECTRAL MODELS OF THE EARTH'S ATMOSPHERIC EVOLUTION

Here we describe the modeling undertaken to generate synthetic spectra of the Earth at different stages in its evolution. Section 2.1 describes the radiative transfer model used to generate the spectra, and its validation. Section 2.2 describes the spectral signature of surfaces, clouds, and land plants. Section 2.3 describes the rationale, development and calculation of the model atmospheres used as input to the radiative transfer model.

## 2.1 MODELS DESCRIPTION AND VALIDATION

Our radiative transfer model is based on a model that was originally developed to model the Earth's atmospheric spectra (Traub & Stier 1978) and has since been used extensively for analyzing high resolution Fourier Transform spectra from ongoing stratospheric balloon-based observations to study the photochemistry and transport of the Earth's atmosphere (for example, Jucks et al. 1998). Our line-by-line radiative transfer code has also been used for numerous full planetary disk modeling studies, both for theoretical studies (Des Marais et al. 2000; Traub and Jucks 2002) and fitting observed Earthshine spectra (Woolf et al. 2002; Turnbull et al. 2006).

We model Earth's reflected and thermal emission spectra using only it's spectroscopically most significant molecules ($H_2O$, $O_3$, $O_2$, $CH_4$, $CO_2$, and $N_2O$). We divide the atmosphere into 28 thin layers from 0 to 100 km altitude. The spectrum is calculated at very high spectral resolution, with several points per line width, where the line shapes and widths are computed using Doppler and pressure broadening on a line-by-line basis, for each layer in the model atmosphere. We use a simple geometrical model in which the spherical Earth is modeled with a plane parallel atmosphere and a single angle of incidence and reflection (visible), or emission (thermal infrared). This angle is selected to give the best analytical approximation to the integrated-Earth air mass factor of two for a nominal illumination (quadrature); the zenith angle of this ray is 60 deg. The overall high-resolution spectrum is calculated, and smeared to lower resolution. For reference and further explanation concerning the code, the reader is referred to our calculation of a complete set of molecular constituent spectra, for a wide range of mixing ratios, for the present Earth pressure-temperature profile, for the visible and thermal infrared, in Des Marais et al. (2002).

We assume that the light paths through the atmosphere can be approximated by 4 parallel streams. All streams traverse the same molecular atmosphere, but each stream reflects (visible), or emits (thermal infrared), from a different lower surface. For example, in the visible and near-infrared, the first stream reflects from the planet's surface at 0 km altitude, the second and third stream reflects from a cloud layer with a top at an adjustable height (here 1 km and 6 km for present Earth), and the fourth stream reflects from a cloud at a high altitude (here 12 km for present Earth). Likewise, in the thermal infrared, these streams originate, rather than reflect, from the corresponding surfaces. The altitude of each cloud layer is adapted according to the height of the tropopause for each epoch presented. On Earth we adopt an overall 60% cloud coverage. The relative proportions of each stream were set to be consistent with the Earthshine data discussed earlier (Woolf et al. 2002; Turnbull et al. 2006). We assume in our model that that fraction and nature of clouds has not changed appreciably over time. We do not model hazes in the atmosphere because) our empirical evidence is that in the present atmosphere the optically-thick model cloud layers effectively mimic the effect of optically-thin real haze layers. Likewise, in the early atmosphere the haze that is expected from CH4 photolysis is probably weak because the $CH_4/CO_2$ ratio does not exceed unity in our models, thus organic haze is not predicted to form (Pavlov et al. 2003).

The reflection or emission spectra of all 4 streams should be considered as only first approximations to reality. For the cumulus and cirrus spectra, the literature contains a variety of cloud spectra that presumably represent a variety of real clouds, but that also may represent a variety of experimental conditions. Parameters that are desirable to know but which are often not specified include the following: the angle of incidence, the angle of reflection, the cloud thickness, the effects of multiple scattering, the degree of homogeneity of the cloud, the frequency of occurrence of a given type of cloud, and the altitude at which a given cloud might form. An additional uncertainty in published cloud spectra is the degree to which the transmission spectrum of the free atmosphere has been removed from the cloud data.

## 2.2 SPECTRAL SIGNATURES OF SURFACES, CLOUDS, LAND AND PLANTS

On a cloud-free planet with surface features like the Earth, the diurnal flux variation in the visible caused by different surface features rotating in and out of view could be high, assuming hemispheric inhomogeneity (Ford 2001; Seager & Ford 2002). When the planet is only partially illuminated, compared to a fully-illuminated case, a more concentrated signal from surface features could be detected as they rotate in and out of view. Most surface features like ice or sand show very small or very smooth continuous reflectivity changes with wavelength. We demonstrate this in Fig. 1, where each panel shows the spectral signature of one specific surface in the visible and near-infrared modeled for our current Earth atmosphere. No clouds were added in Fig. 1 to emphasize the detailed surface features. The top (red) smooth lines show the albedo of the surfaces without an overlaying atmosphere (albedo data from ASTER 1999, USGS 2003), the black lines show the modeled atmosphere of current Earth assuming one particular surface denoted in each panel. These individual panels are coadded percentage-wise to model a realistic Earth surface.

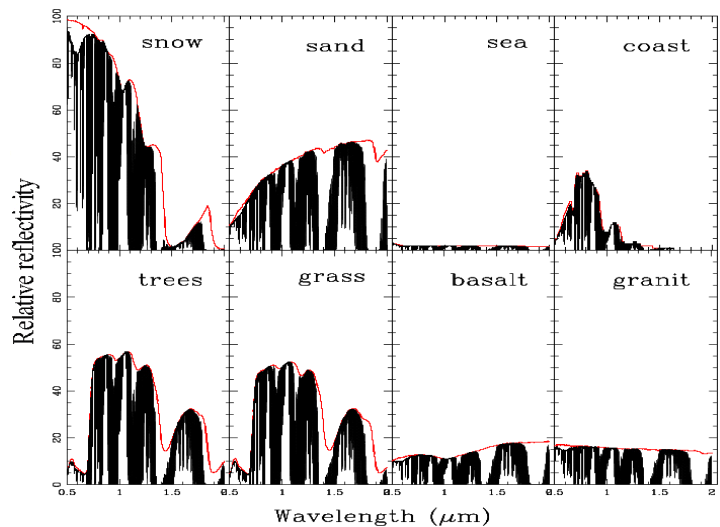

**Fig. 1: Signal of a present-day Earth atmosphere considering different surface features assuming a clear atmosphere without clouds.**

In our model we assign 70% of the planetary surface as ocean, 2% as coast, and 28% as land. The land surface consists of 30% grass, 30% trees, 9% granite, 9% basalt, 15% snow and 7% sand for present-day Earth. The land surface changes to 35% basalt, 40% granite, 15% snow and 10% sand for a younger Earth, where no land vegetation is present (A. Knoll, private communication). Fig 1. shows that an ocean planet will be more difficult to detect than a similar size snow-covered planet due to its low albedo in the visible, assuming no clouds or a similar cloud coverage for both planets. In the thermal infrared its lower surface temperature will reduce the emitted signal. Assuming a planet without clouds seems likely to be an unrealistic approach. We include the spectra for a cloudless and an Earth-like cloud fraction in this paper to put our work in context with different groups that concentrate on models without clouds.

Clouds are an important component of exoplanet spectra because their reflection is high and relatively flat with wavelength. Clouds hide the atmospheric molecular species below them, weakening the spectral lines in both the thermal infrared and visible (see Discussion). In the thermal infrared, clouds emit at temperatures that are generally colder than the surface, while in the visible the clouds themselves have different spectrally-dependent albedos that further influence the overall shape of the spectrum. Fig. 2 shows the normalized visible reflection and thermal infrared spectral emission of the Earth for 3 cloud conditions. The cloud altitudes and weights in the average spectra in Fig. 2 are: low cloud, 1 km, 40%; medium cloud, 6 km, 40%; and high cloud, 12 km, 20% for today's Earth. The other curves represent complete overcast conditions with clouds at low, medium, and high altitudes. As is apparent from Fig. 2, clouds tend to reduce the relative depths, full widths, and equivalent widths of spectral features. For our final models the spectra from the surfaces and clouds are coadded with the indicated weights.

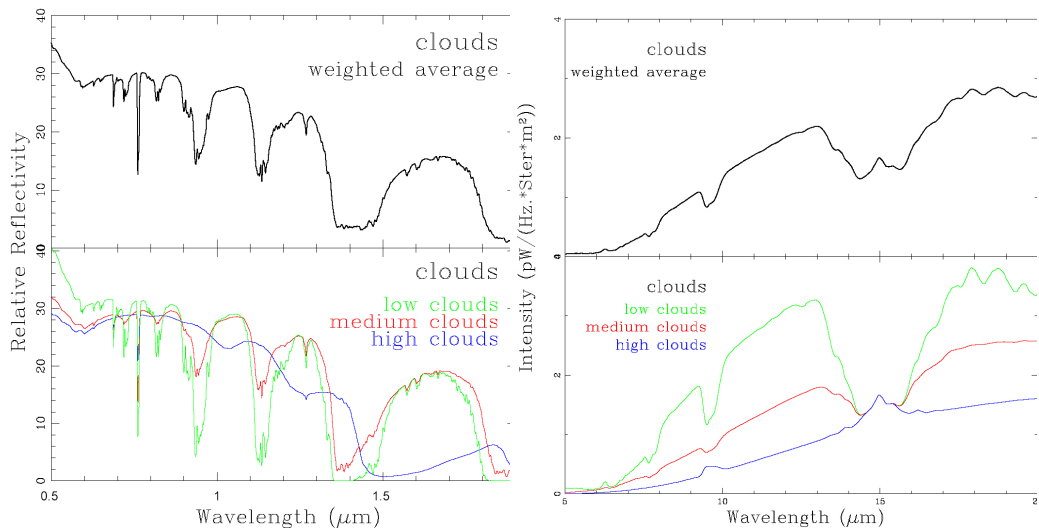

**Fig. 2:** (lower) Spectra of present-day Earth with (green) 100% cumulus cloud coverage at 1 km and (red) 100% cumulus cloud coverage at 6 km and (blue) 100% cirrus cloud coverage at 12 km, and, (upper) spectra of a mixture of clouds resembling the present Earth for the visible (left) and thermal infrared (right).

Our theoretical model has been validated by comparison to observed spectra. Fig. 3 shows observations and model fits to spectra of the Earth in 3 wavelength ranges. In each case the constituent gas spectra in a clear atmosphere are shown in the bottom panel, for reference. Fig. 3 (left) is the visible Earthshine spectrum (Woolf et al. 2002). Fig. 3 (center) is the near-infrared Earthshine spectrum (Turnbull et al. 2006); and Fig. 3 (right) is the thermal infrared spectrum of Earth as measured by a spectrometer enroute to Mars (Christensen & Pearl 1997). The data are shown in black and our model in red.

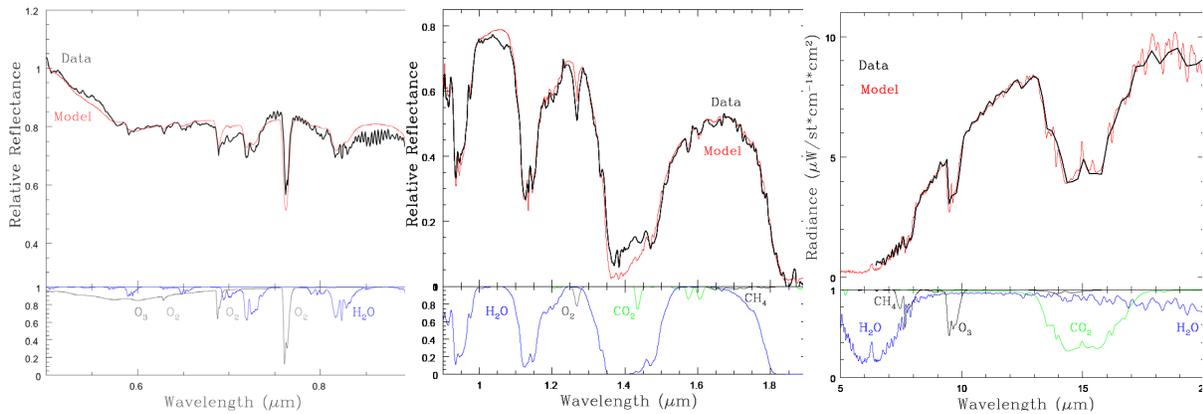

**Fig. 3: Observed reflectivity spectrum in the visible (Woolf et al. 2002), near-infrared (Turnbull et al. 2006) and emission spectrum in the infrared (Christensen & Pearl 1997) of the integrated Earth, as determined from Earthshine and space respectively. The data is shown in black and our model in red. The reflectivity scale is arbitrary.**

The model provides an excellent fit to these observed data sets. In the following section we extend these results back in time, and present the spectra of Earth over geological times, calculated using the radiative transfer model just described, plus appropriate model atmospheres.

## 2.3 RATIONALE, DEVELOPMENT AND CALCULATION OF MODEL ATMOSPHERES OVER GEOLOGICAL TIME

Our models of the history of the Earth's atmosphere are based on studies in the literature. These studies include Kasting and Catling (2003), Kasting (2004), Pavlov et al. (2000), and Segura et al. (2003). These models include the concentration profiles of the spectrally most significant atmospheric molecules as well the temperature profile, while considering the solar input and molecular oxygen concentration for specific scenarios. The surface pressure is set to 1 bar. These studies are used to determine the inputs to our radiative transfer model through geological time.

The base of our model is the tropospheric abundances of five key gases as a function of geological time adapted from Kasting (2004; private communication, 2006), and shown in Fig. 4 with numerical values in Table 1. The indicated abundances are notional and reflect only long term trends, however they are sufficient for our present purpose. To estimate the effects of increased levels of $CH_4$ and $CO_2$ levels in early atmospheres on the temperature profile and the $H_2O$ mixing ratio, we use atmosphere profiles calculated by Segura et al. (2003) as a starting point and linearly add on the effects of an increase in $CH_4$ and $CO_2$ for each layer as calculated by Pavlov et al. (2000) for an atmosphere without oxygen.

Kasting and Catling (2003) and Kasting (2004) established a guideline for the Earth's atmosphere evolution. Pavlov et al. (2000) calculated Earth atmosphere profiles for a young sun with 0.71 times the present luminosity for no oxygen but different levels of $CH_4$ and $CO_2$. Pavlov's model provides mixing ratio and temperature profiles up to 30 km. We adapt a constant temperature structure between 30 km and the exosphere on the basis that there are no stratospheric-heating species that could act as an analog to today's stratospheric ozone layer (see also Walker 1977). Segura et al. (2003) calculated Earth atmosphere profiles for varying levels of oxygen in a present day atmosphere and current solar output.

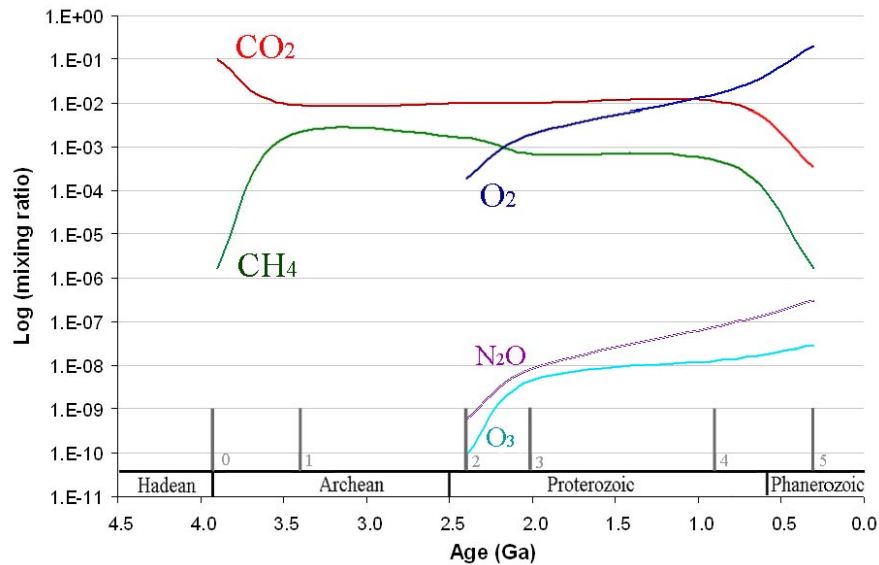

**Fig. 4: Schematic evolution of abundances of key atmospheric species over geological time (based on Kasting (2004))**

**Table 1: Evolution of surface abundances over geological time (based on Kasting (2004))**

| Epoch | Age (Ga) | $CO_2$ (mixing ratio) | $CH_4$ (mixing ratio) | $O_2$ (mixing ratio) | $O_3$ (mixing ratio) | $N_2O$ (mixing ratio) |
|---|---|---|---|---|---|---|
| 0 | 3.9 | 1.00E-01 | 1.65E-06 | 0 | 0 | 0 |
| 1 | 3.5 | 1.00E-02 | 1.65E-03 | 0 | 0 | 0 |
| 2 | 2.4 | 1.00E-02 | 7.07E-03 | 2.10E-04 | 8.47E-11 | 5.71E-10 |
| 3 | 2.0 | 1.00E-02 | 1.65E-03 | 2.10E-03 | 4.24E-09 | 8.37E-09 |
| 4 | 0.8 | 1.00E-02 | 4.15E-04 | 2.10E-02 | 1.36E-08 | 9.15E-08 |
| 5 | 0.3 | 3.65E-04 | 1.65E-06 | 2.10E-01 | 3.00E-08 | 3.00E-07 |

From the schematic evolution of abundances shown in Fig. 4, we chose 6 epochs that reflect significant states in the chemical composition of the atmosphere. These epochs and the corresponding ages and tropospheric mixing ratios are listed in Table 1. Following standard practice, we use the term mixing ratio to mean the fractional number density of a species. With the addition of molecules with no spectral opacity (e.g., $N_2$), the sum of mixing ratios at any altitude will be unity.

Our model atmosphere ranges from a $CO_2$ rich atmosphere (3.9 Ga = epoch 0) to a $CO_2/CH_4$-rich atmosphere (epoch 3) to a present- day atmosphere (epoch 5 = present-day Earth) (see Fig. 4 and Table 1) (Ga = $10^9$ years ago). In this paper we focus on long-lived periods in the Earth's history and we ignore relatively short-term events such as glaciation events (the "snowball Earth") and their warm counterparts (the "hothouse Earth"); we will address the spectra of these and other events in detail in a later paper.

Fig. 5 through Fig. 8 show the temperature and chemical mixing ratio profiles for the six epochs. Profiles for $O_2$ and $CO_2$ are not shown because they are assumed to be well-mixed (i.e., constant in mixing ratio from 0 to 100 km). Note that the temperature profiles show a distinctive shape for the different epochs. The $N_2O$ increase from epoch 2 to epoch 5 occurs because of increased shielding of solar UV radiation by $O_3$ see Fig. 5 (Segura et al. 2003). We assume negligible $O_2$ and $N_2O$ concentration for the calculations of early Earth's atmosphere (epoch 0 to epoch 1), thus those epochs are not shown in Fig. 7 and Fig. 8. For an overview of oxygen evolution over geological times see Holland (2006).

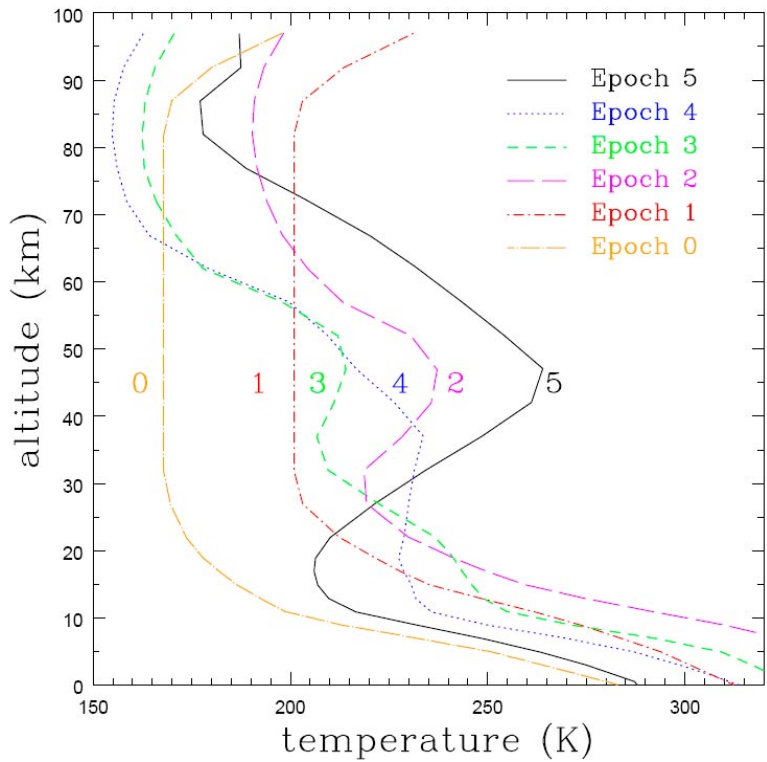

**Fig. 5: Temperature profile for an Earth-like planet over its evolution. The surface was cool in epochs 0 and 1, hot to warm in epochs 2, 3, and 4, and cool again in epoch 5.**

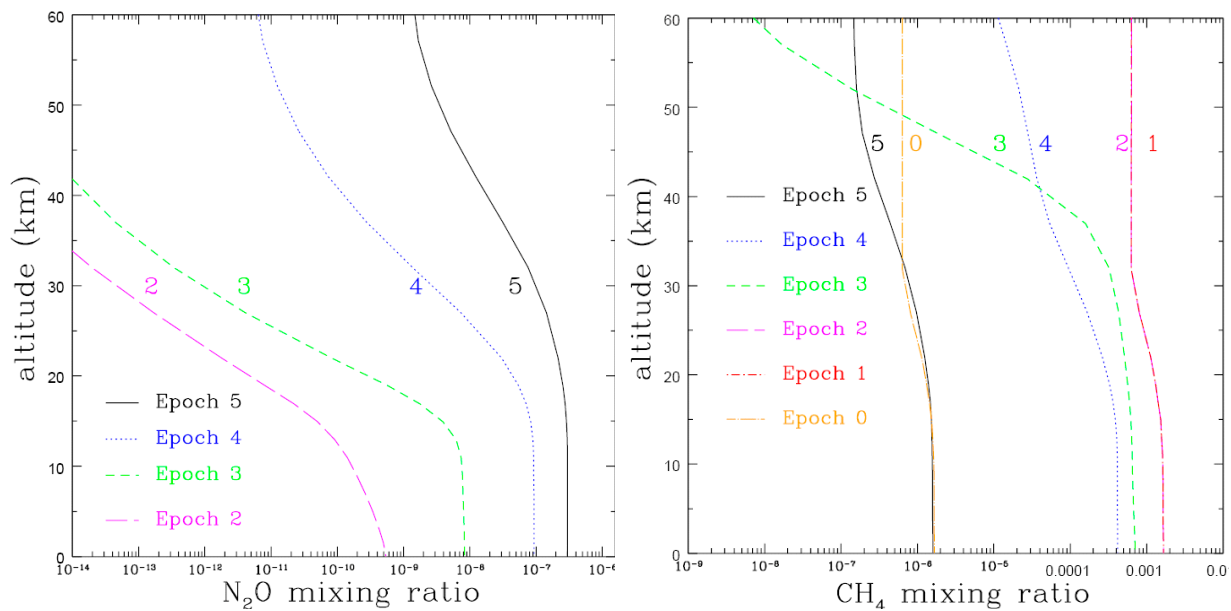

**Fig. 6: CH$_4$ and H$_2$O mixing ratio for an Earth-like planet over its evolution. Methane had a low abundance at the beginning (epoch 0) to a maximum (epoch 2) , then fell off to its present level.**

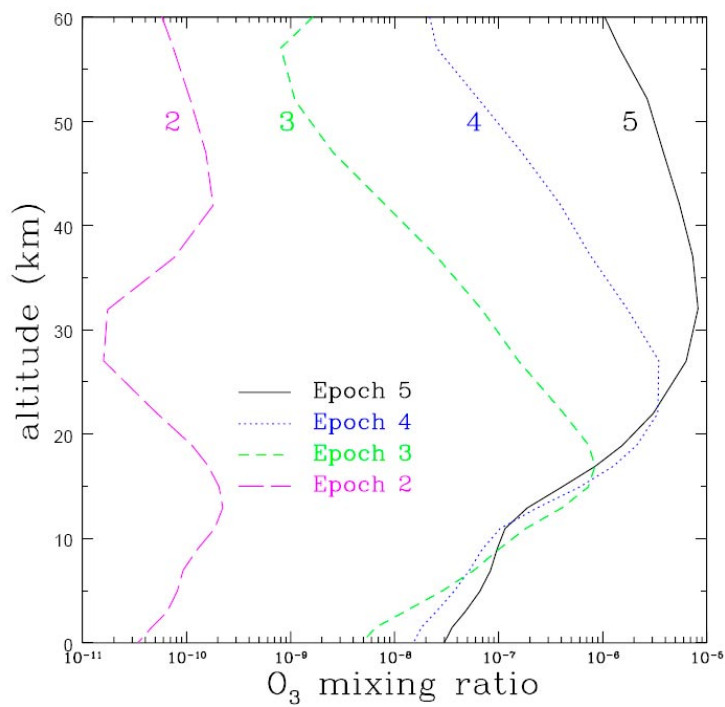

**Fig. 7: O$_3$ number density for an Earth-like planet over its evolution. Ozone had a negligible abundance at the beginning (epoch 0), then steadily rose in abundance from epoch 2 to epoch 5 (present).**

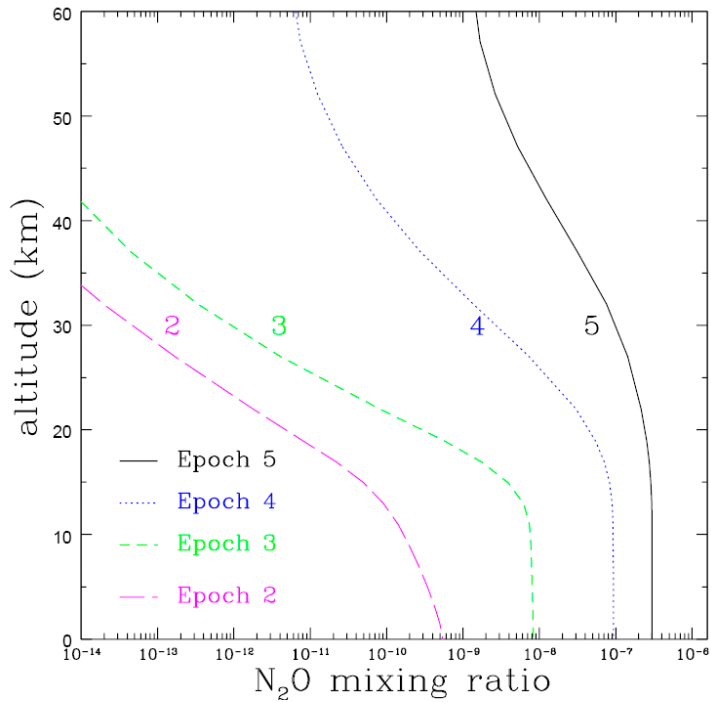

**Fig. 8: N₂O mixing ratio for an Earth-like planet over its evolution. Nitrous oxide increased in time, roughly in parallel with O₂ and O₃, being negligible in epochs 0 and 1, then rising rapidly from epochs 2 to 5 (present).**

To set our model in context with the overall Earth evolution we sketch out the conditions on Earth prior to epoch 0. The Earth formed about 4.5 billion years ago. The primitive atmosphere was formed by the release of volatiles from the interior, and/or volatiles delivered during the late bombardment period. Standard models of solar evolution predict that the sun was 30% less luminous at 4.6 Ga and has brightened monotonically since that time. Because of the faint young sun, Earth's mean surface temperature should have been below the freezing point of seawater before about 2.0 Ga, if the Bond albedo was similar to today's, and even if there was a similar greenhouse contributions to the temperature (see e.g., Sagan & Muller 1972; Kasting 1993, 1997; 2002). However, geological records tell us that liquid water was present by at least 3.5 Ga (Sagan & Muller 1972) and probably 4.0 Ga (Mojizsis et al. 1996). The oldest zircon crystals are as old as 4.4 Ga (Wilde et al. 2001) suggesting that liquid water formed even earlier than 4 Ga. This argues for a higher abundance of greenhouse gases in the early atmosphere to keep the surface temperature above the freezing point of water (see also e.g. Kasting et al. 2004).

## 2.3.0. Epoch 0 (3.9 Ga)

Epoch 0 is centered at about 3.9 Ga. The atmosphere was most likely dominated by carbon dioxide that originated from volcanoes, with nitrogen being the most abundant gas and trace amounts of methane. Therefore for our input model for this epoch we use 10% $CO_2$, current amounts of $CH_4$, and no $O_2$, $O_3$, or $N_2O$ in the atmosphere.

## 2.3.1. Epoch 1 (3.5 Ga)

Epoch 1 (about 3.5 Ga) reflects the decrease of carbon dioxide and the rise of methane in the early atmosphere. Between epoch 0 and epoch 1, a significant amount of $CO_2$ must have been removed from the atmosphere, most likely by the burial of carbon into carbonate rocks, though the process is still debated. The major influence of methane on the atmosphere may have begun almost as soon as life originated more than 3.5 billion years ago (Kasting & Siefert 2002; Ueno et al. 2006; Canfield 2006). Methanogens are believed to have produced methane levels roughly 1000 times that to today. $CH_4$ could have been quite abundant in an anoxic atmosphere. $CH_4$ has only a 10-year residence time today because it reacts with the hydroxyl radical, OH, and $O^1D$. In an anoxic atmosphere, OH and $O^1D$ would have been much less abundant and $CH_4$ would have been destroyed mainly by photolysis at Ly α wavelengths (121.6 nm). Under such conditions, its residence time should have been more like 10,000 years (Pavlov et al. 2001). A biogenic $CH_4$ source comparable to the modern flux of 535 Tg $CH_4$/year, which produces an atmospheric $CH_4$ concentration of 1.6 ppm (parts per million) today, could have generated over 1000 ppm of $CH_4$ in the distant past (Kasting & Siefert 2002). This is enough to have had a major warming effect on climate (Pavlov et al. 2003). The atmosphere in epoch 1 consists mainly of $N_2$ and $CO_2$ with $CH_4$ becoming a major component. Therefore for our model we use 1% $CO_2$, 0.2% $CH_4$, and no $O_2$, $O_3$, or $N_2O$ in the atmosphere.

## 2.3.2. Epoch 2 (2.4 Ga)

Epoch 2 (about 2.4 Ga) reflects a maximum level of methane, a constant carbon dioxide concentration and a small trace of oxygen in the early atmosphere. Most methanogens grow best at temperatures above 40 °C; some even prefer at least 85 °C. Those that thrive at higher temperatures grow faster, so as the intensifying greenhouse effect raised temperatures at the planet's surface, more of these faster-growing, heat-loving specialists would have survived (Kasting 2004). As they made up a larger proportion of the methanogen population, more methane molecules would have accumulated in the atmosphere, making the surface temperature still warmer—in fact, hotter than today's climate, despite the dimmer sun. The factor that limited the $CH_4$ abundance was likely the production of organic haze, which is predicted to form when the atmospheric $CH_4/CO_2$ ratio exceeds unity (Pavlov et al. 2003). This haze would have created an "anti–greenhouse effect," which would have lowered surface temperatures and made life less comfortable for the predominately thermophilic methanogens (Cooney 1975), thus reducing the amount of $CH_4$ in the atmosphere. In our models we keep the $CH_4/CO_2$ ratio below unity. Primitive cyanobacteria are believed to have produced oxygen. At some point in Earth's history organisms discovered how to perform photosynthesis. The oxygen produced from this reaction is responsible for most of the $O_2$ in Earth's present atmosphere (Kasting & Catling 2003). Oxygen is toxic to methanogens. The atmosphere in epoch 2 consists mainly of $N_2$, about equal amounts of $CO_2$ and $CH_4$ and small amounts of oxygen. Therefore for our model, we use 1% $CO_2$, 0.7% $CH_4$, 0.02% $O_2$, and trace amounts of $O_3$ and $N_2O$ in the atmosphere.

## 2.3.3. Epoch 3 (2.0 Ga)

Epoch 3 is centered at about 2.0 Ga. It reflects the rise of oxygen and decrease of methane in the early atmosphere. At some time between epochs 2 and 3, the Earth's atmosphere underwent a dramatic change (see Kasting 2004 for an overview). From a variety of geological evidence, we know that significant concentrations of free $O_2$ began to appear in the atmosphere (see e.g Cloud 1972; Farquhar et al. 2000, 2001; Kasting 2002; Pavlov et al. 2001; Ono et al. 2003). This marked a sharp transition from basically anoxic to $O_2$-rich conditions. As remarked by Raymond and Segre (2006), Falkowski (2006) "the introduction of $O_2$ into an anaerobic biosphere around 2.2 billion years ago must have represented a cataclysm in the history of life". Between epoch 2 and epoch 3, the abundance of oxygen rises in the atmosphere and Earth goes through a major glaciation event that is thought to be related to the drop in methane concentration in the atmosphere due to the rise of oxygen. Epoch 3 reflects a time after most of the reduced minerals were oxidized, and atmospheric oxygen started to accumulate in the atmosphere. The atmosphere in epoch 3 consists mainly of $N_2$, constant $CO_2$, and about equal amounts of $CH_4$ and $O_2$. Oxygen accumulates in the atmosphere while $CH_4$ decreases. Therefore for our model, we use 1% $CO_2$, 0.4% $CH_4$, 0.2% $O_2$, and increasing trace amounts of $O_3$ and $N_2O$ in the atmosphere.

Global ice ages occurred at least three times in the Proterozoic era, first at 2.3 Ga and again at 0.75 and 0.6 Ga. The circumstances surrounding these glaciations were long un-explained, but the methane hypothesis provides compelling answers (Kasting & Siefert 2002). The rise in atmospheric $O_2$ corresponds precisely with Earth's first well-documented glaciation (Prasad & Roscoe 1996), suggesting that the glaciation was triggered by the accompanying decrease in atmospheric $CH_4$.

## 2.3.4. Epoch 4 (0.8 Ga)

Epoch 4 (about 0.8 Ga) reflects a further rise of oxygen (Kennedy et al. 2006) and further decrease of carbon dioxide in the atmosphere. Different schemes have been suggested to quantify the rise of oxygen and the evolution of life by anchoring the points in time to fossil finds (see e.g. Owen 1980; Schopf 1993; Ehrenfreund & Charnley 2003). There are still many open questions. Carbon isotope data suggest that production of $O_2$ was occurring at rates comparable to today (Kump et al. 2001). Therefore the sinks for $O_2$ must have been larger. Canfield (1998), Canfield, Habicht & Thamdrup (2000), and others (e.g. Anbar & Knoll 2002) have argued that the deep oceans remained anoxic (and sulfidic) during most of the Proterozoic. This suggests that at least until 0.6-0.8 Ga atmospheric $O_2$ levels remained significantly lower than today. We use this model for our atmosphere calculations. Epoch 4 reflects an increase in oxygen by a factor of 10 and a consequential decrease in $CO_2$ and $CH_4$. The atmosphere in epoch 4 consists mainly of $N_2$, 1% of $CO_2$, 0.04% $CH_4$, 2% $O_2$, and further increases in the trace species $O_3$ and $N_2O$.

## 2.3.5. Epoch 5 (0.3 Ga)

Epoch 5 (about 0.3 Ga to present-day Earth) reflects the present-day Earth's atmosphere, and also the influence of vegetation on our climate (see also Tinetti et al. 2006a; Kaltenegger et al. 2006; Meadows 2006; Seager & Ford 2002; Woolf et al. 2002; Des Marais et al. 2002). The atmosphere consists mainly of $N_2$, with 0.0365 % $CO_2$ and 21 % $O_2$ as the second most abundant species followed by present day trace amounts of $CH_4$, $O_3$, and $N_2O$. We use this atmosphere profile to model our balloon and Earthshine measurements. It shows an excellent fit to the data, as discussed in section 2.2.

# 3. SPECTRAL SIGNATURE OF EARTH OVER GEOLOGICAL TIME

## 3.0 SPECTRA

Using the geological model atmospheres and radiative-transfer methods sketched in Sec. 2, we calculate the Earth's spectra for 6 geological epochs, shown in Fig. 9 (visible to near-infrared) and Fig. 10 (thermal infrared). Major observable molecular species ($H_2O$, $O_3$, $O_2$, $CH_4$, $CO_2$, and $N_2O$) are labeled. The dark lines show a resolution of 70 in the visible and 20 in the thermal infrared, as proposed for the TPF-C and Darwin/TPF-I mission, respectively. Fig. 9 and Fig. 10 show that the changes in atmospheric signatures are clearly visible in both the visible and thermal infrared over Earth's evolution.

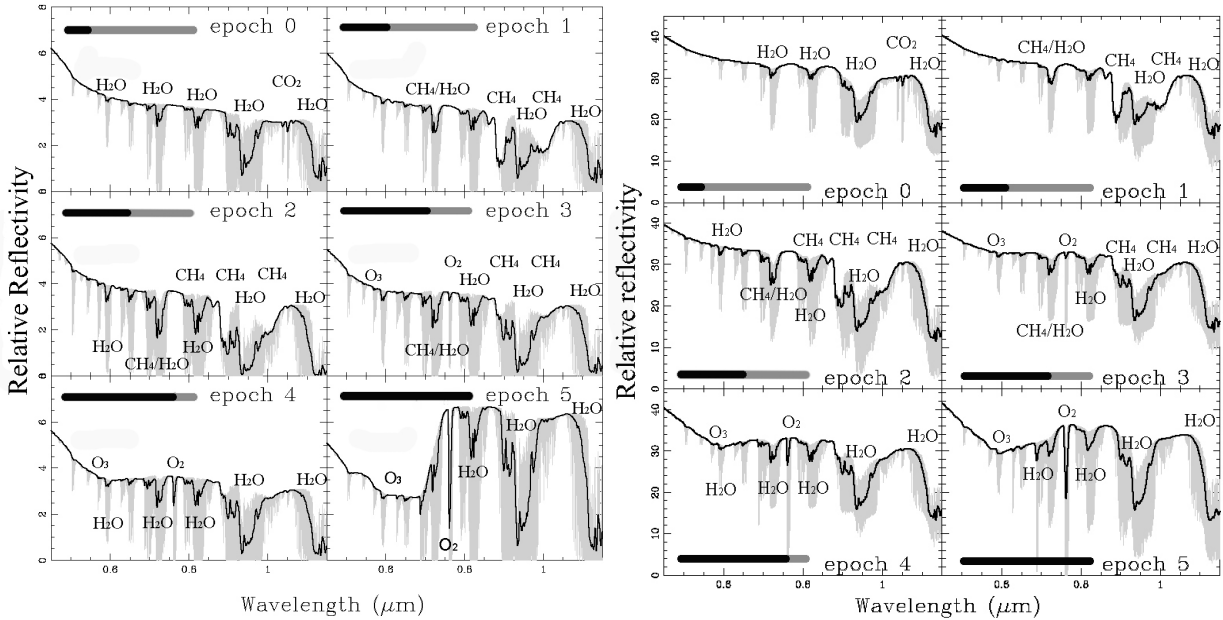

**Fig. 9: Visible and near infrared spectra of an Earth-like planet for 6 geological epochs. Spectra without clouds (left) and with clouds (right) are shown; note the different scales. The spectral features change considerably as the planet evolves from a CO₂ rich (epoch 0) to a CO₂/CH₄-rich atmosphere (epoch 3) to a present-day atmosphere (epoch 5) in the visible. The black lines show spectral resolution of 70, comparable to the proposed TPF-C mission concept.**

In the visible to near-infrared one can see increasingly strong $H_2O$ bands at 0.73 μm, 0.82 μm, 0.95 μm, and 1.14 μm. These can be seen throughout the Earth's evolution. The strongest $O_2$ feature is the saturated Fraunhofer A-band at 0.76 μm that can be clearly seen from epoch 3 to epoch 5. It is still relatively strong for significantly smaller mixing ratios than present Earth's as seen in epoch 3 and epoch 4. A weaker feature at 0.69 μm can not be seen with low resolution. $O_3$ has a broad feature, the Chappius band, which appears as a broad triangular dip in the middle of the visible spectrum from about 0.45 μm to 0.74 μm that can be seen from epoch 3 to epoch 5. The feature is very broad and shallow thus in epoch 3 this feature is only marginally detectable. Methane at present terrestrial abundance (1.65 ppm) has no significant visible absorption features but at high abundance as seen in epoch 1 to epoch 3, it has strong visible bands at 0.88 μm, and 1.04 μm, readily detectable in early Earth's history. $CO_2$ has negligible visible features at present abundance, but in a high $CO_2$-atmosphere of 10% seen in the early evolution stage of epoch 0, the weak 1.06 μm band could be observed. Clouds, as shown in the right panel of Fig 9, hide the atmospheric molecular species below them, essentially weakening the spectral lines. In the visible the clouds themselves have different spectrally dependent albedos that further influence the overall shape of the spectrum (see Fig. 9 right panel vs. left panel) (King et al. 1990). In epoch 5 we can clearly see the red edge of land plants in our model. As land coverage occurred about 0.44 Ga ago, this red edge can not be observed before epoch 5.

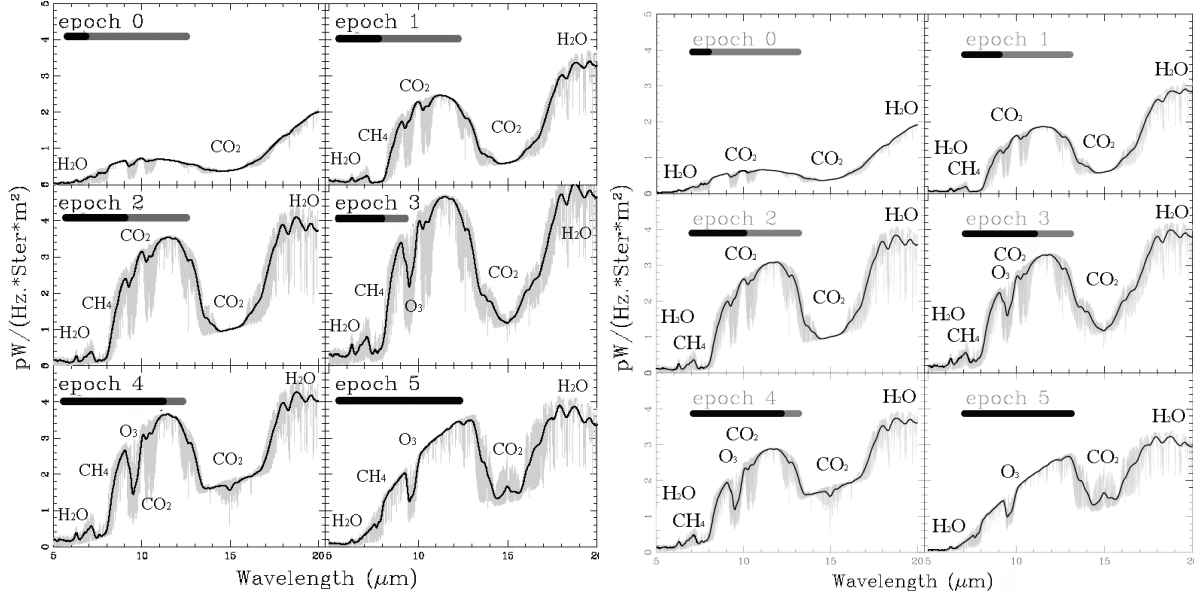

**Fig. 10: Same as Fig. 9 for the thermal infrared with a resolution of 20.**

In the thermal infrared the 9.6 μm $O_3$ band can be seen from epoch 3 to epoch 5. It is highly saturated and is thus an excellent qualitative but a poor quantitative indicator for the existence of even traces of the parent species ($O_2$). Ozone is a very nonlinear indicator of $O_2$ because the ozone column depth changes slowly as $O_2$ increases from 0.01 present atmosphere level (PAL) to 1 PAL. $CH_4$ is not readily identified in our present day atmosphere using low resolution spectroscopy (epoch 5), but the methane feature at 7.66 μm in the thermal infrared is easily detectable for epoch 1 to epoch 4. There are three weak $N_2O$ features in the infrared at 7.75 μm, 8.52 μm and 16.89 μm. These features are strongly overlapped by $CH_4$, $CO_2$ and $H_2O$, so it is unlikely to become a prime target for the first generation of space-based missions searching for exoplanets that will work with low resolution, but it is a good target for follow up missions because it is an excellent biomarker.

In the thermal infrared clouds, as shown in the right panel of Fig. 10, emit at temperatures that are generally colder than the surface. They hide the atmospheric molecular species below them, essentially weakening the spectral lines, just as for the visible region.

In the thermal infrared the classical signatures of biological activity are the combined detection of 9.6 μm $O_3$ band, the 15 μm $CO_2$ band and the 6.3 μm $H_2O$ band or its rotational band that extends from 12 μm out into the microwave region (Selsis & Despois 2002). These can be detected from epoch 3 to epoch 5.

## 3.1. EQUIVALENT WIDTHS

This section provides quantitative information on the strength of spectral features on an Earth-like exoplanet over geological time. We calculate the equivalent width of each chemical signature in the visible, near-infrared and thermal infrared. Following standard practice, the total absorption in the feature is expressed in terms of equivalent width, (i.e., the spectral width of an equal area of a rectangular line with zero residual intensity and unity continuum).

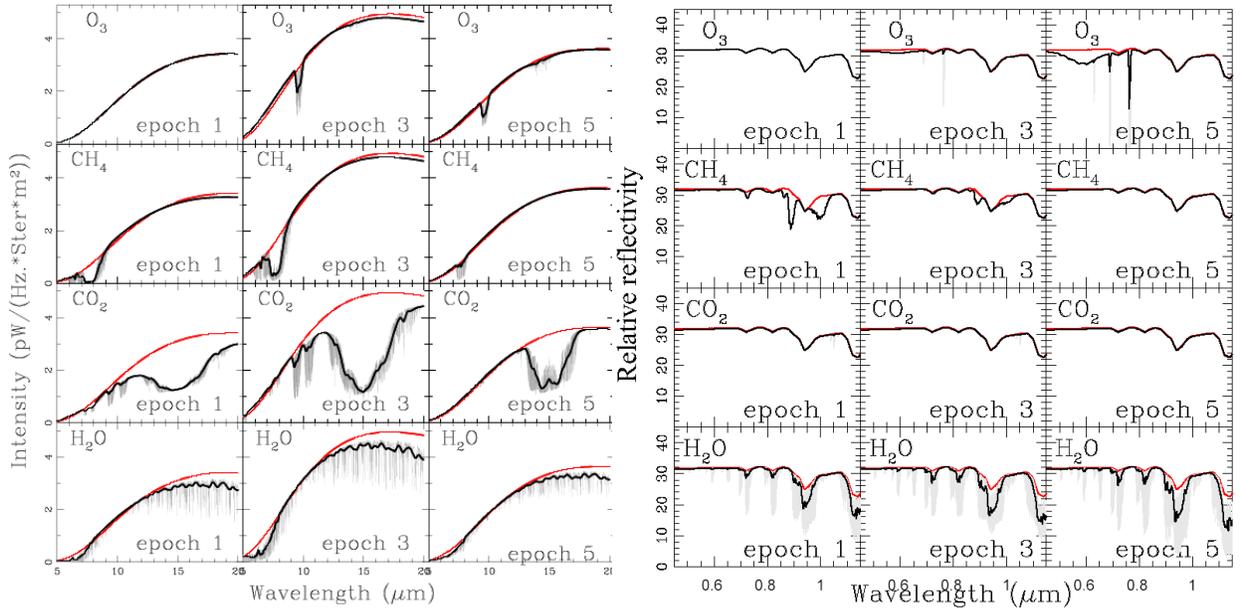

**Fig. 11: Evolution of individual chemical signatures over geological time for a partially cloud covered Earth, in the visible (left) and thermal infrared (right) from a $CO_2$ rich (3.9 Ga = epoch 1) to a $CO_2/CH_4$-rich atmosphere (epoch 3) to a present-day atmosphere (epoch 5). The black lines show spectral resolution of 70 for the visible and 20 for the thermal infrared, the top (red) line shows the reference spectra with no chemical absorptions.**

In order to emphasize the changes in the spectral features of each molecule over the epochs, Fig. 11 shows the evolution of our model spectra for individual chemical species over geological time, for a partially cloudy Earth. The red line in each graph shows the reference spectrum without any atmospheric chemical feature present. Since many of the spectral features from the different molecules overlap, a more descriptive picture is one where the difference between the calculated spectra with all molecules and spectra with all molecules except the molecule of interest are plotted. These are shown for the cloud free models in the left panels of Fig.12 and Fig. 13. Similarly, the present day cloud models are plotted in the right panels of Fig. 12 and Fig. 13. Clouds hide the atmospheric molecular species below them, essentially weakening the spectral lines and are a very important component of exoplanet spectra.

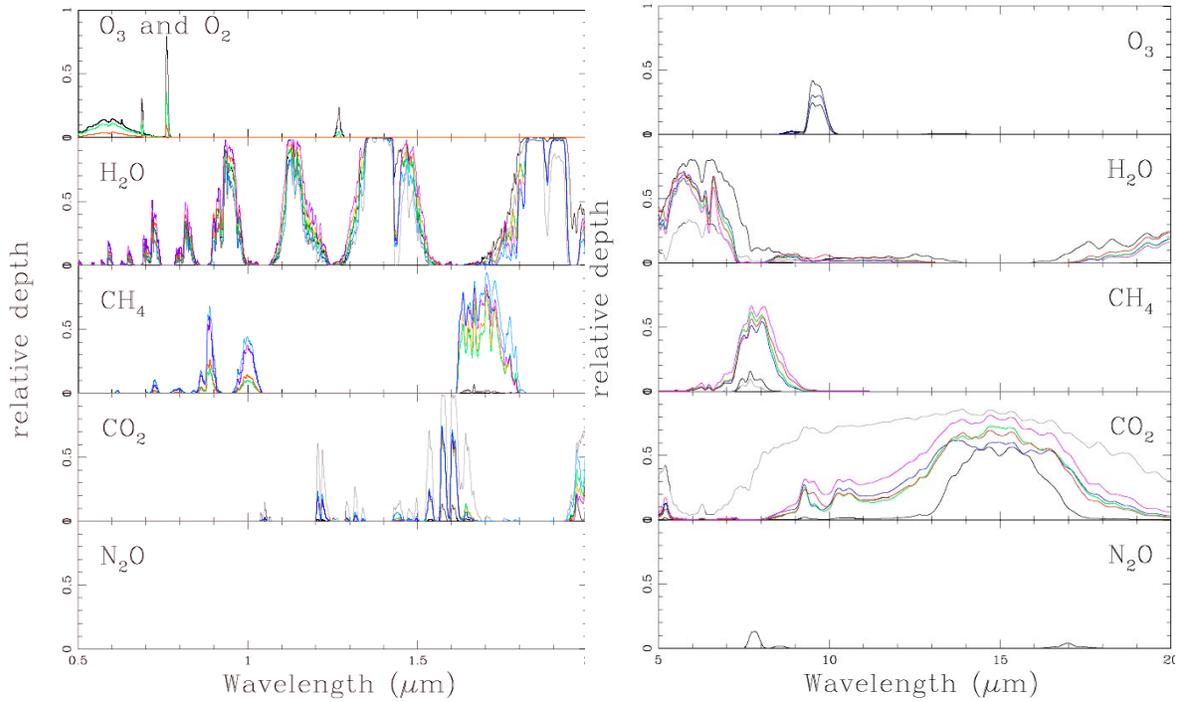

**Fig. 12: The relative depth of spectral features, i.e., (continuum-spectrum)/continuum, is shown here, for the visible (left), and thermal infrared (right), for individual atmospheric species and for a cloud free atmosphere. The colors denote epochs: grey (0), magenta (1), red (2), green (3), blue (4), black (5).**

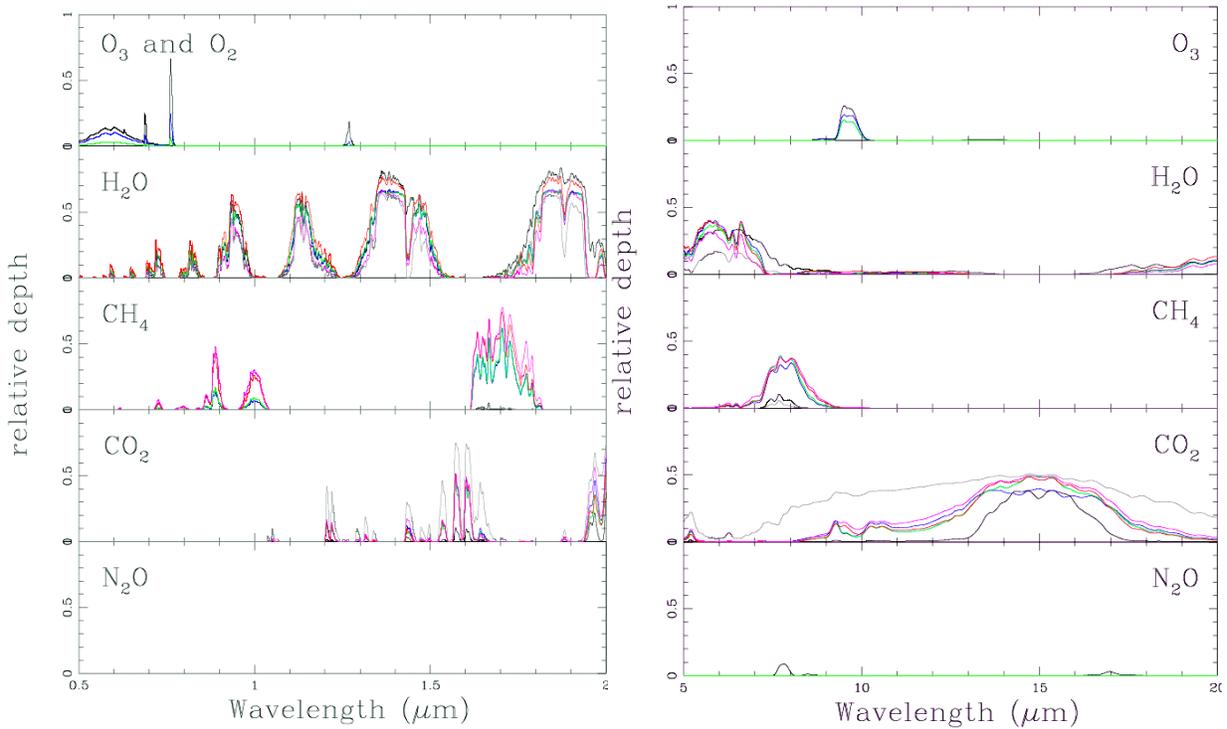

**Fig. 13: Same as Fig. 12 but for atmospheres with a present-day cloud distribution.**

We use the curves in Fig. 12 and Fig. 13 to numerically calculate the equivalent widths of each spectral feature in a straight forward fashion by integrating the area of each feature in these curves. Table 2, 4 and 6 show the results for a cloud free model while Table 3, 5 and 7 show the results for a planet with Earth-like cloud coverage. The equivalent widths don't necessarily map directly with the abundances for several reasons. As stated above, some of the spectral features overlap, especially with the water vapor bands. Variations in water vapor due to changes in the atmospheric temperature structure between the epochs mean that each equivalent width calculation is unique. Therefore, significant non-linear structure is noticeable in the equivalent width of each structure over time.

**Table 2: Equivalent width (nm) of the main spectral features of atmospheric compounds in the visible vs. geological age with no clouds.**

| Feature | $\lambda$ ($\mu$m) | 0.3 Ga | 0.8 Ga | 2.0 Ga | 2.4 Ga | 3.5 Ga | 3.9 Ga |
|---|---|---|---|---|---|---|---|
| $H_2O$ | 0.95 | 47.56 | 48.69 | 55.70 | 62.92 | 37.00 | 35.39 |
| $H_2O$ | 0.82 | 12.48 | 12.31 | 14.81 | 17.82 | 7.85 | 8.34 |
| $H_2O$ | 0.73 | 6.88 | 7.03 | 8.89 | 11.73 | 4.62 | 4.35 |
| $CH_4$ | 1.04 | - | 4.16 | 5.67 | 15.23 | 19.40 | - |
| $CH_4$ | 0.89 | - | 3.80 | 5.16 | 12.85 | 16.07 | - |
| $CO_2$ | 1.05 | 0.01 | 0.19 | 0.18 | 0.22 | 0.22 | 1.39 |
| $O_2$ | 0.76 | 5.83 | 2.55 | 0.71 | - | - | - |
| $O_2$ | 0.69 | 1.67 | 1.03 | 0.38 | - | - | - |
| $O_3$ | 0.59 | 19.28 | 13.79 | 5.16 | - | - | - |

**Table 3: Equivalent width (nm) of the main spectral features of atmospheric compounds in the visible vs. geological age with Earth clouds.**

| Feature | $\lambda$ ($\mu$m) | 0.3 Ga | 0.8 Ga | 2.0 Ga | 2.4 Ga | 3.5 Ga | 3.9 Ga |
|---|---|---|---|---|---|---|---|
| $H_2O$ | 0.95 | 29.23 | 28.76 | 30.18 | 37.25 | 19.03 | 17.97 |
| $H_2O$ | 0.82 | 7.38 | 7.07 | 7.64 | 10.02 | 3.88 | 4.09 |
| $H_2O$ | 0.73 | 4.00 | 3.99 | 4.49 | 6.48 | 2.23 | 2.08 |
| $CH_4$ | 1.04 | - | 2.95 | 3.72 | 10.80 | 13.40 | - |
| $CH_4$ | 0.89 | - | 2.72 | 3.40 | 9.31 | 11.31 | - |
| $CO_2$ | 1.05 | 0.01 | 0.13 | 0.11 | 0.15 | 0.15 | 0.88 |
| $O_2$ | 0.76 | 4.90 | 1.98 | 0.49 | - | - | - |
| $O_2$ | 0.69 | 1.37 | 0.98 | 0.31 | - | - | - |
| $O_3$ | 0.59 | 18.66 | 12.88 | 4.12 | - | - | - |

**Table 4: Equivalent width (nm) of the main spectral features of atmospheric compounds over geological times in the near-infrared with no clouds.**

| Feature | $\lambda$ ($\mu$m) | 0.3 Ga | 0.8 Ga | 2.0 Ga | 2.4 Ga | 3.5 Ga | 3.9 Ga |
|---|---|---|---|---|---|---|---|
| $H_2O$ | 1.86 | 200.27 | 155.02 | 155.29 | 146.74 | 137.84 | 139.29 |
| $H_2O$ | 1.41 | 171.19 | 162.44 | 173.23 | 182.89 | 146.14 | 127.40 |
| $H_2O$ | 1.14 | 55.41 | 56.10 | 62.92 | 71.51 | 45.97 | 41.93 |
| $CH_4$ | 1.66 | 1.97 | 68.58 | 74.02 | 92.08 | 112.76 | 1.70 |
| $CO_2$ | 1.57 | 3.72 | 27.26 | 26.57 | 27.43 | 27.15 | 70.84 |
| $CO_2$ | 1.21 | 0.25 | 3.09 | 2.87 | 2.97 | 3.55 | 11.08 |
| $O_2$ | 1.27 | 2.49 | 0.49 | 0.05 | - | - | - |

**Table 5: Equivalent width (nm) of the main spectral features of atmospheric compounds over geological times in the near-infrared with Earth clouds.**

| Feature | λ (μm) | 0.3 Ga | 0.8 Ga | 2.0 Ga | 2.4 Ga | 3.5 Ga | 3.9 Ga |
|---------|--------|--------|--------|--------|--------|--------|--------|
| $H_2O$ | 1.86 | 143.03 | 100.78 | 99.04 | 105.59 | 86.95 | 84.05 |
| $H_2O$ | 1.41 | 117.80 | 101.95 | 103.85 | 120.18 | 85.13 | 73.56 |
| $H_2O$ | 1.14 | 34.71 | 33.52 | 34.71 | 43.09 | 24.08 | 21.76 |
| $CH_4$ | 1.66 | 1.57 | 53.93 | 54.84 | 77.84 | 89.25 | 1.14 |
| $CO_2$ | 1.57 | 2.82 | 19.16 | 17.12 | 18.90 | 18.16 | 47.30 |
| $CO_2$ | 1.21 | 0.20 | 2.24 | 1.93 | 2.15 | 2.42 | 7.32 |
| $O_2$ | 1.27 | 1.92 | 0.34 | 0.03 | - | - | - |

**Table 6: Equivalent width (μm) of the main spectral features of atmospheric compounds over geological times in the thermal infrared with no clouds.**

| Feature | λ (μm) | 0.3 Ga | 0.8 Ga | 2.0 Ga | 2.4 Ga | 3.5 Ga | 3.9 Ga |
|---------|--------|--------|--------|--------|--------|--------|--------|
| $H_2O$ | 19.51 | 0.66 | 0.28 | 0.28 | 0.40 | 0.22 | 0.01 |
| $H_2O$ | 6.50 | 1.65 | 1.15 | 1.06 | 1.14 | 0.96 | 0.52 |
| $CO_2$ | 15.06 | 1.79 | 4.69 | 4.50 | 4.43 | 6.39 | 14.67 |
| $CH_4$ | 7.81 | 0.09 | 0.67 | 0.75 | 0.87 | 1.03 | 0.09 |
| $O_3$ | 9.61 | 0.22 | 0.17 | 0.13 | - | - | - |
| $N_2O$ | 16.96 | 0.03 | 0.0010 | 0.00005 | - | - | - |
| $N_2O$ | 7.78 | 0.05 | 0.0006 | 0.00003 | | | |

**Table 7: Equivalent width (μm) of the main spectral features of atmospheric compounds over geological times in the thermal infrared with Earth clouds.**

| Feature | λ (μm) | 0.3 Ga | 0.8 Ga | 2.0 Ga | 2.4 Ga | 3.5 Ga | 3.9 Ga |
|---------|--------|--------|--------|--------|--------|--------|--------|
| $H_2O$ | 19.51 | 0.27 | 0.16 | 0.15 | 0.22 | 0.10 | 0.01 |
| $H_2O$ | 6.50 | 0.68 | 0.66 | 0.59 | 0.64 | 0.46 | 0.26 |
| $CO_2$ | 15.06 | 1.19 | 2.89 | 2.78 | 2.84 | 3.53 | 5.73 |
| $CH_4$ | 7.81 | 0.05 | 0.36 | 0.41 | 0.44 | 0.49 | 0.02 |
| $O_3$ | 9.61 | 0.14 | 0.10 | 0.08 | - | - | - |
| $N_2O$ | 16.96 | 0.02 | 0.0006 | 0.00003 | | | |
| $N_2O$ | 7.78 | 0.03 | 0.0004 | 0.00002 | - | - | - |

## 3.2. RESOLUTION

To detect a spectral feature with optimum signal to noise requires that the FWHM of the spectrometer should be approximately equal to the FWHM of the spectral features. The spectral resolution ($\lambda/\Delta\lambda$) needed for optimal detection of each changing spectral feature over geological time is given in Table 8 through 10 for the cloudy atmosphere. Here $\lambda$ is the central wavelength of a feature, and $\Delta\lambda$ is the FWHM of the feature as determined directly from Figs. 12 and 13. Since many of the features, even after smoothing, have significant structure and are asymmetric, we determine the furthest out points on either side the features' maximum that is half the intensity of the maximum to define the FWHM. Because of the spectral structure, and the effects of saturation and overlap, this can result in significant changes in the calculated resolution for different concentrations. An example of this is demonstrated by the $CO_2$ at 1.54 microns. This feature has been defined by the 4 lobes seen in the difference spectra. When $CO_2$ abundances are high, the width encompasses all 4 lobes, while when they are low, the width only encompasses the middle two lobes, resulting in a higher required resolution. These numbers are relevant for the design of missions like Darwin and TPF, which currently concentrate on biomarkers in our present day atmosphere.

**Table 8: Resolution ($\lambda/\Delta\lambda$) needed to match the main spectral features of atmospheric compounds over geological time in the visible with clouds.**

| Feature | $\lambda$ ($\mu$m) | 0.3 Ga | 0.8 Ga | 2.0 Ga | 2.4 Ga | 3.5 Ga | 3.9 Ga |
|---------|---------|--------|--------|--------|--------|--------|--------|
| $CH_4$ | 1.04 | - | 26 | 27 | 27 | 24 | - |
| $CH_4$ | 0.89 | - | 54 | 54 | 53 | 48 | - |
| $CO_2$ | 1.05 | 184 | 202 | 198 | 202 | 210 | 202 |
| $H_2O$ | 0.95 | 25 | 5 | 5 | 5 | 28 | 27 |
| $H_2O$ | 0.82 | 9 | 9 | 8 | 8 | 9 | 9 |
| $H_2O$ | 0.73 | 47 | 46 | 45 | 45 | 49 | 50 |
| $O_2$ | 0.76 | 125 | 136 | 131 | - | - | - |
| $O_2$ | 0.69 | 143 | 128 | 128 | - | - | - |
| $O_3$ | 0.59 | 5 | 4 | 5 | - | - | - |

In the visible $O_3$ has a very broad feature that can be detected at 0.59 $\mu$m and a resolution of 5. $H_2O$ can be detected at 0.82 $\mu$m at a resolution of 9 throughout all Earth's evolution. $CH_4$ can be detected from epoch 1 to epoch 4 at 1.04 $\mu$m with a resolution of 24 to 27. $O_2$ can be detected in epoch 3 to epoch 5 at 0.76 $\mu$m at a resolution of 125 to 136. $CO_2$ can be detected at 1.05$\mu$m at a resolution of 184 to 202.

**Table 9: Resolution ($\lambda/\Delta\lambda$) needed to match the main spectral features of atmospheric compounds over geological time in the near-infrared with clouds.**

| Feature | $\lambda$ ($\mu$m) | 0.3 | 0.8 | 2.0 | 2.4 | 3.5 | 3.9 |
|---------|---------|-----|-----|-----|-----|-----|-----|
| $CH_4$ | 1.66 | 325 | 15 | 15 | 14 | 10 | 325 |
| $H_2O$ | 1.86 | 11 | 13 | 13 | 13 | 13 | 13 |
| $H_2O$ | 1.41 | 9 | 8 | 8 | 8 | 9 | 11 |
| $H_2O$ | 1.14 | 23 | 22 | 21 | 20 | 25 | 26 |
| $O_2$ | 1.27 | 165 | 240 | 244 | - | - | - |
| $CO_2$ | 1.57 | 37 | 34 | 33 | 34 | 34 | 13 |
| $CO_2$ | 1.21 | 63 | 58 | 58 | 57 | 58 | 55 |

In the near-infrared waveband we can detect $H_2O$ throughout all Earth's evolution at 1.41$\mu$m and 1.14$\mu$m with a resolution of 8 to 13. $CO_2$ can be detected from throughout Earth's evolution at 1.57$\mu$m at a resolution of 13 to 37. $O_2$ can be detected in epoch 3 to epoch 5 at 1.27$\mu$m at a resolution of 165 to 244. $CH_4$ can be detected at 1.66$\mu$m at a resolution of 10 to 325.

**Table 10: Resolution ($\lambda/\Delta\lambda$) needed to match the main spectral features of atmospheric compounds over geological times in the thermal infrared with clouds.**

| Feature | $\lambda$ ($\mu$m) | 0.3 | 0.8 | 2 | 2.4 | 3.5 | 3.9 |
|---------|---------|-----|-----|---|-----|-----|-----|
| $CO_2$ | 15.06 | 5 | 3 | 4 | 4 | 3 | 1 |
| $CH_4$ | 7.81 | 12 | 8 | 8 | 7 | 6 | 23 |
| $H_2O$ | 19.51 | 7 | 19 | 19 | 19 | 20 | 25 |
| $H_2O$ | 6.5 | 4 | 3 | 4 | 4 | 4 | 4 |
| $O_3$ | 9.61 | 19 | 18 | 18 | - | - | - |
| $N_2O$ | 16.96 | 44 | 17 | 17 | - | - | - |
| $N_2O$ | 7.78 | 23 | 20 | 20 | - | - | - |

In the thermal infrared, the optimal detection of $H_2O$ is at 6.5μm at a resolution of 3 to 4 and at 19.51μm at a resolution of 7 to 25 throughout all Earth's evolution. Note that only part of the 19.51μm feature is detected due to the wavelength cutoff at 20μm. $CH_4$ can be detected at 7.81μm at a resolution of 6 to 23. $O_3$ can be detected in epoch 3 to epoch 5 at 9.61μm at a resolution of 18 to 19. $N_2O$ can only be detected in epoch 3 to epoch 5 at 7.78μm at a resolution of 20 to 23.

The graphical interpretation for Table 2 through 10 can be found in Appendix A.1 and A.2.

## 4 DISCUSSION

In this section we comment briefly on the relationship between the results presented in this paper and the larger issue of detecting signs of habitability or actual signs of life on an exoplanet. We also comment on spectral changes that are expected compared to the models in this paper, if an exoplanet has a dramatically different type of cloud cover. Finally we comment on the possible change in the visible spectrum if land plants have developed differently on an exoplanet than on Earth.

The work presented here is a comprehensive study of the evolution of the Earth's atmosphere over geological time. It improves on preliminary studies by exploring more epochs, treats the surface more realistically, includes the effect of clouds on the spectra and quantifies the spectral resolution required to detect habitability and biosignatures as a function of time.

The spectrum of the Earth has not been static throughout the past 4.5 Ga. This is due to the variations in the molecular abundances, the temperature structure, and the surface morphology over time. At about 2.3 Ga oxygen and ozone became abundant, affecting the atmospheric absorption component of the spectrum. At about 2 Ga, a green phytoplankton signal developed in the oceans and at about 0.44 Ga, an extensive land plant cover followed, generating the red chlorophyll edge in the reflection spectrum. We find that the composition of the surface (especially in the visible), the atmospheric composition, and temperature-pressure profile can all have a significant influence on the detectabilty of a signal (see Fig 9 and Fig 10). Note that we assume that the cloud cover over these epochs is the same as the cloud cover today, however changes in this distribution could significantly change the overall spectra in both wavelength regions.

Our search for signs of life is based on the assumption that extraterrestrial life shares fundamental characteristics with life on Earth, in that it requires liquid water as a solvent and has a carbon-based chemistry (Owen 1980; Des Marais et al. 2002). Therefore we assume that extraterrestrial life is similar to life on Earth in its use of the same input and output gases, that it exists out of thermodynamic equilibrium, and that it has analogs to bacteria, plants, and animals on Earth (Lovelock 1975). Biomarkers are detectable species whose presence at significant abundance requires a biological origin (see also Des Marais et al. 2002 for an overview). They are the chemical ingredients necessary for biosynthesis (e.g., $O_2$ and $CH_4$) or are products of biosynthesis (e.g., complex organic molecules and cells, but also $O_2$ and $CH_4$).

As signs of life in themselves $H_2O$ and $CO_2$ are secondary in importance because although they are raw materials for life, they are not unambiguous indicators of its presence, as both can be seen for a long time in Earth's atmosphere ($H_2O$ from epoch 0 through epoch 5 in both wavelength ranges, $CO_2$ from epoch 0 to epoch 5 in the thermal infrared) before life developed. Taken together with molecular oxygen (that can be seen from epoch 3 through epoch 5 in the visible as $O_2$ and in both the visible and thermal infrared as $O_3$), abundant $CH_4$ (visible from epoch 1 to epoch 4) can indicate biological processes (see also Lovelock 1975, Segura et al. 2003). Depending on the degree of oxidation of a planet's crust and upper mantle non-biological mechanisms can also produce large amounts of $CH_4$ under certain circumstances.

Oxygen and $N_2O$ are two very promising bio-indicators. Because oxygen is a chemically reactive gas, it follows that reduced gases and oxygen have to be produced concurrently to be detectable in the atmosphere, as they react rapidly with each other. The oxygen and ozone absorption features in the visible and thermal infrared respectively could have been used to indicate the presence of photosynthetic biological activity on Earth anytime during the past 50% of the age of the solar system, while the chlorophyll red-edge reflection feature evolved during the most recent 10% of the age of the solar system. $N_2O$ is a very interesting chemical because it is produced in abundance by life but only in trace amounts by natural processes. There are no $N_2O$ features in the visible and three weak $N_2O$ features in the thermal infrared at 7.75 μm and 8.52 μm, and 16.89 μm. For present-day Earth one needs a resolution of 23 and 44 respectively to detect $N_2O$. Spectral features of $N_2O$ also become more apparent in atmospheres with less $H_2O$ vapor. Methane, detectable with low resolution from epoch 1 to epoch 4, and nitrous oxide, detectable in epoch 5, have features nearly overlapping in the 7 μm region, and additionally both lie in the red wing of the 6 μm water band. Thus $N_2O$ is unlikely to become a prime target for the first generation of space-based missions searching for exoplanets, but it is an excellent target for follow up missions. There are other molecules that could, under some circumstances, act as excellent biomarkers, e.g., the manufactured chloro-fluorocarbons ($CCl_2F_2$ and $CCl_3F$) in our current atmosphere in the thermal infrared waveband, but their abundances are too low to be spectroscopically observed at low resolution.

We also present the spectral resolution required to detect habitability and biosignatures as a function of time for an Earth-like planet. These numbers are relevant for the design of missions like Darwin and TPF, which currently concentrate on biomarkers in our present day atmosphere. To detect and characterize an Earth-like exoplanet the spectral resolution required is dependent on the evolution stage and should be taken into account in the design to expand the scope of detectable habitable planets.

Clouds are a very important component of exoplanet spectra as their high reflection is relatively flat with wavelength. Clouds hide the atmospheric molecular species below them, essentially weakening the spectral lines in both the thermal infrared and the visible. Fig. 14 shows all three cloud layers and their influence on the spectra over the whole visible to near-infrared spectrum and the change they add onto the model spectra of our Earth over geological time scales.

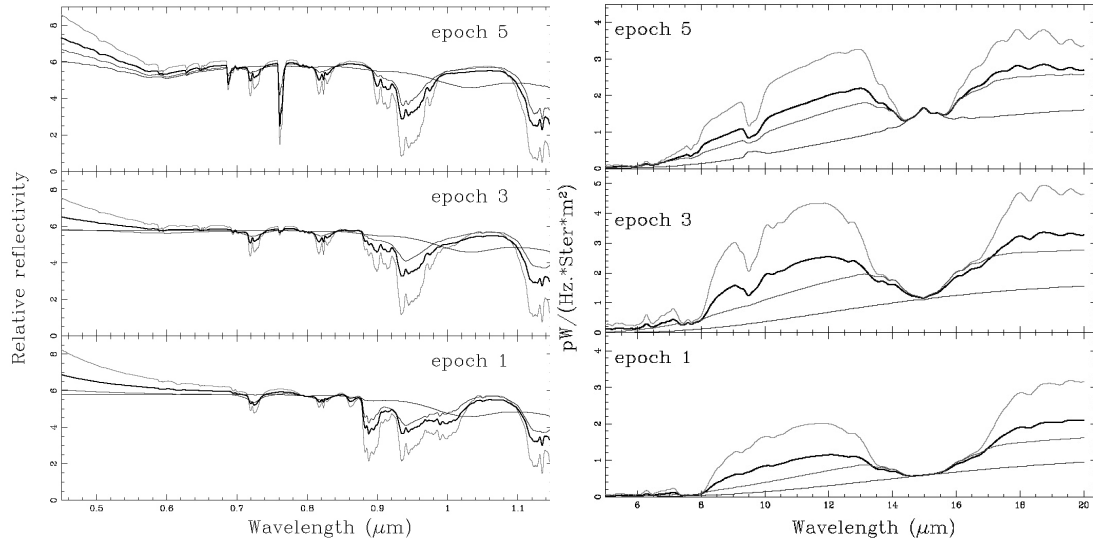

**Fig. 14:** (left) Visible reflection and (right) infrared thermal emission spectral models for three cloud conditions for an Earth-like planet over its evolution from a $CO_2$ rich atmosphere (epoch 1) to a $CO_2/CH_4$-rich atmosphere (epoch 3) to present day atmosphere (epoch 5). The averaged cloud spectrum is shown as the bold black line.

A dramatic case of surface biomarkers on Earth is the red edge signature from photosynthetic plants at about 720 nm. About 0.44 Ga, an extensive land plant cover developed, generating the red chlorophyll edge in the reflection spectrum. Grass cover could have been abundant from 0.65 Ga (Piperno & Sues 2005), however it may not have been extensive.

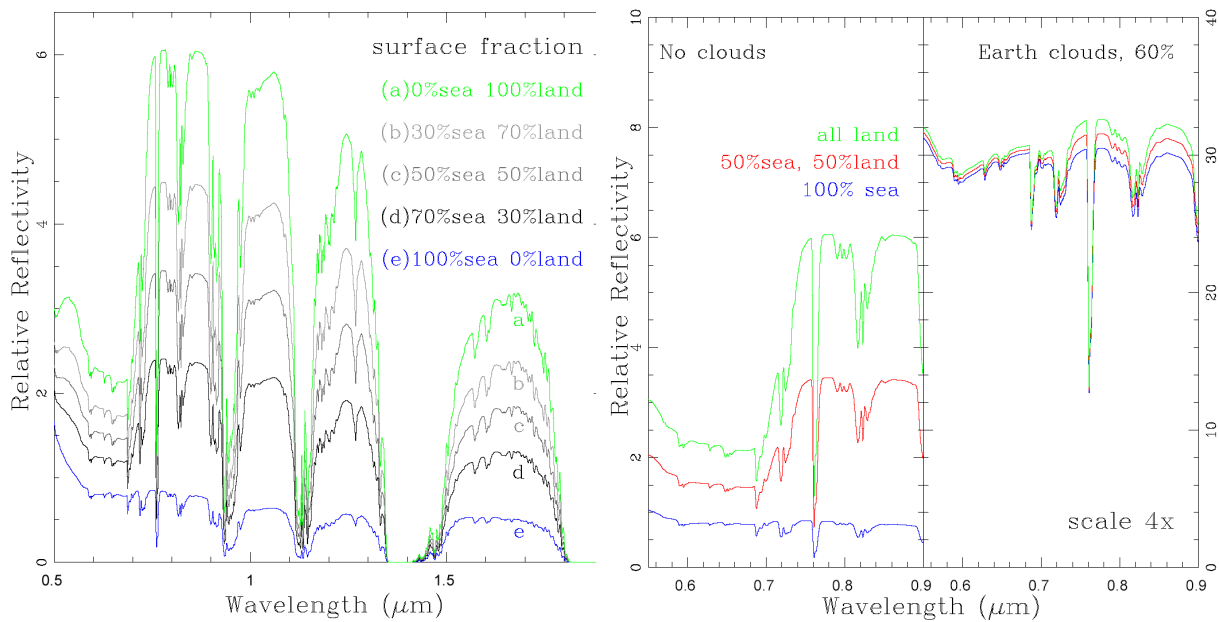

**Fig. 15:** (left) Signal of different surface fractions of sea and on land for present-day Earth atmosphere, for several contributions of different kind of surfaces assuming a clear atmosphere without clouds. (left) Spectra of present-day Earth without (left) and with (right) clouds for a disk averaged view of the ocean and landmasses. Note that the low albedo of the ocean reduces the overall flux and the different scale of the left plot.

Photosynthetic plants have developed strong infrared reflection, possibly as a defense against overheating and chlorophyll degradation. The primary molecules that absorb the energy and convert it to drive photosynthesis ($H_2O$ and $CO_2$ into sugars and $O_2$) are chlorophyll A (0.450 µm) and B (0.680 µm). Several groups have measured the integrated Earth spectrum via the technique of Earthshine, using sunlight reflected from the non-illuminated, or "dark", side of the moon. Earthshine measurements have shown that detection of Earth's vegetation-red edge (VRE) is feasible but made difficult owing to its broad, essentially featureless spectrum and cloud coverage (e.g. Montanes-Rodriguez et al. 2006, Turnbull et al. 2005). For an Earth-like planet we show the influence of vegetation on a cloud free spectrum in Fig. 15. One sees that the red edge feature becomes predominant with an increase in percentage of the surface that is vegetation covered.  The sea has a very low reflectivity and dilutes the red edge and overall albedo.

Trying to identify such weak, continuum-like features at unknown wavelengths in an exoplanet spectrum requires models for different planetary conditions. Our knowledge of the reflectivity of different surface components on Earth, like deserts, ocean and ice, help in assigning the VRE of the Earthshine spectrum to terrestrial vegetation (see Fig. 15). Earth's hemispherically integrated vegetation red-edge signature is weak, but planets with different rotation rates, obliquities, land-ocean fraction, and continental arrangement may have lower cloud-cover and higher vegetated fraction (see e.g. Seager & Ford 2002).

The exact wavelength and strength of the spectroscopic "red edge" depends on the plant species and environment. Averaged over a spatially unresolved hemisphere of Earth, the additional reflectivity of this spectral feature is a few percent. The main diluting factors include forest canopy architecture, soil characteristics, non-continuous coverage of vegetation across Earth's surface, and presence of clouds which prevent a view of the surface. Keeping in mind that the chances are unknown that another planet has developed the exact same vegetation as Earth, finding the same signatures might be unlikely. Modeling of different planet spectra, including the possibility of a different type of photosynthesis (Tinetti et al. 2006) is essential to be able to interpret the detections. The signatures of a different type of photosynthesis may be difficult to verify through remote observations as being of biological origin.

To illustrate the dilution of the red edge, Fig. 15 shows the spectra for present-day Earth without (left) and with (right) clouds for a disk averaged view of the ocean and landmasses. Note that there is a factor of 4 in scale between the plots, reflecting the higher cloud albedo with respect to surface albedos. The clouds reduce the VRE considerably, compared to a cloudless planet, and require high resolution detection to determine the percentage of vegetation. The overall flux in the visible is considerably higher for a cloud covered Earth because of the higher cloud albedo with respect to the solid surface, and especially the albedo of sea water that covers 72% of Earth's surface. Fig. 15 emphasizes that a planet without clouds or snow has a lower overall albedo and thus requires a longer integration time for detection than a planet with clouds.

# 5. CONCLUSIONS

Concentrating on the evolution of our planet we established a model for its atmosphere and detectable biomarkers. Our model shows a considerable change with time for the biosignatures in the Earth's atmosphere. Observations of these features on an exoplanet should be able to place an Earth-like planet with regard to its state of evolution. Knowledge of those features will help to optimize the design of proposed instruments to search for Earth-like planets. If an exoplanet is found with a corresponding spectrum, we will have good evidence for characterizing its evolutionary state, its habitability, and the degree to which it shows signs of life.

We showed the results as they pertain to TPF-I/Darwin and TPF-C. These missions should be able to constrain an Earth-like planet in its evolution. The atmospheric features on an Earth-like planet change considerably over its evolution from a $CO_2$ rich (3.9 Ga = epoch 0) to a $CO_2/CH_4$-rich atmosphere (epoch 3 around 2 Ga) to a present-day atmosphere (epoch 5 = present Earth). The spectral resolution required for optimal detection of habitability and biosignatures as a function of time for an Earth-like planet have been discussed. In the visible for optimum detection a spectrometer should have a resolution of: $H_2O$, 8 to 9 throughout Earth's evolution; $CH_4$, 24 to 27 (epoch 1 to epoch 4); $O_3$, 4 to 5 (epoch 3 to epoch 5); $O_2$, 125 to 136 (epoch 3 to epoch 5); $CO_2$, 184 to 210.

In the near-infrared the resolutions are: $H_2O$, 8 to 11; $CO_2$, 13 to 37 throughout Earth's evolution; $O_2$, 165 to 244 (epoch 3 to epoch 5); $CH_4$, 10 to 325 (epoch 0 to epoch 5). Combining the visible and near-infrared leads to a resolution of 8 to 325 needed to detect $H_2O$, $CO_2$, $O_2$ and $CH_4$. In the thermal infrared the resolutions are: $H_2O$, 3 to 4; $CO_2$, 1 to 5; $CH_4$, 7 to 25 throughout Earth's evolution; $O_3$, 18 to 19 (epoch 3 to epoch 5); $N_2O$, 20 to 23 (epoch 3 to epoch 5). These numbers are relevant to the design of missions like Darwin and TPF, which currently concentrate on biomarkers in our present-day atmosphere. To detect and characterize an Earth-like exoplanet the spectral resolution required is dependent on the evolution stage and should be taken into account in the design to expand the scope of detectable habitable planets. Clues to photosynthetic biological activity have been present in Earth's atmosphere for 2 billion years, since epoch 3. For optimal detection a spectrometer should approximately match the features to be detected, and therefore have a spectral resolution of 5 to 210 in the visible, 8 to 325 in the near-infrared and 3 to 23 in the infrared.

Future work will place special events in Earth's history (e.g., the snowball and hothouse events) into the context of the overall evolution model. incorporate cloud models when they become available, and include a self consistent climate model.

**Acknowledgement:**
Special thanks to Jim Kasting, Antigona Segura and Andy Knoll for stimulating discussion and comments and to the referee Vikki Meadows for very constructive suggestions that clarified the content. This work was sponsored by NASA grant NAG5-13045 and the Navigator Program at JPL.

# SUPPLEMENT MATERIAL

The graphical interpretation of the material in Tables 2 through 10 is shown here.

## S.1 EQUIVALENT WIDTHS

We show here the expected strength of spectral features on an Earth-like exoplanet over geological time (see Fig. 16 through Fig.21). The equivalent width of each chemical signature in the visible, near-infrared and thermal infrared is shown, and in each case for an Earth with no clouds and an Earth with present-day cloud types and coverage.

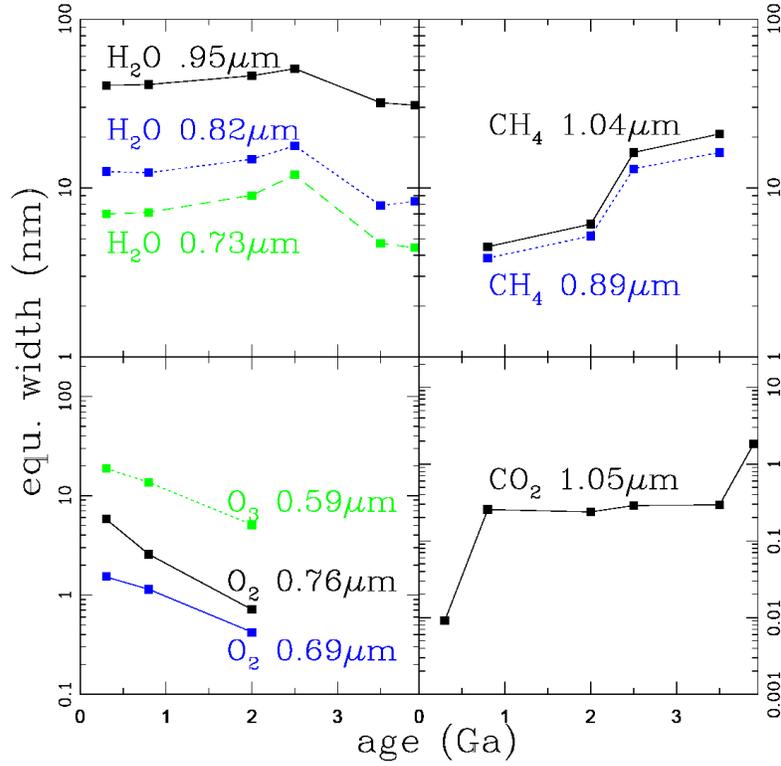

**Fig. 16: Equivalent width (nm) of the main spectral features of atmospheric compounds in the visible vs. geological age with no clouds (see table 2).**

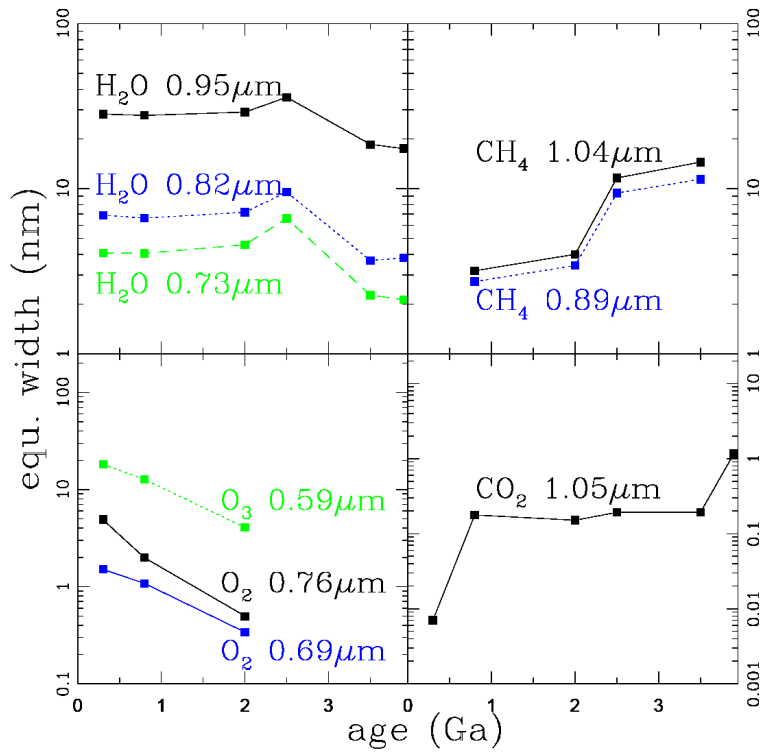

**Fig. 17: Equivalent width (nm) of the main spectral features of atmospheric compounds in the visible vs. geological age with Earth clouds.**

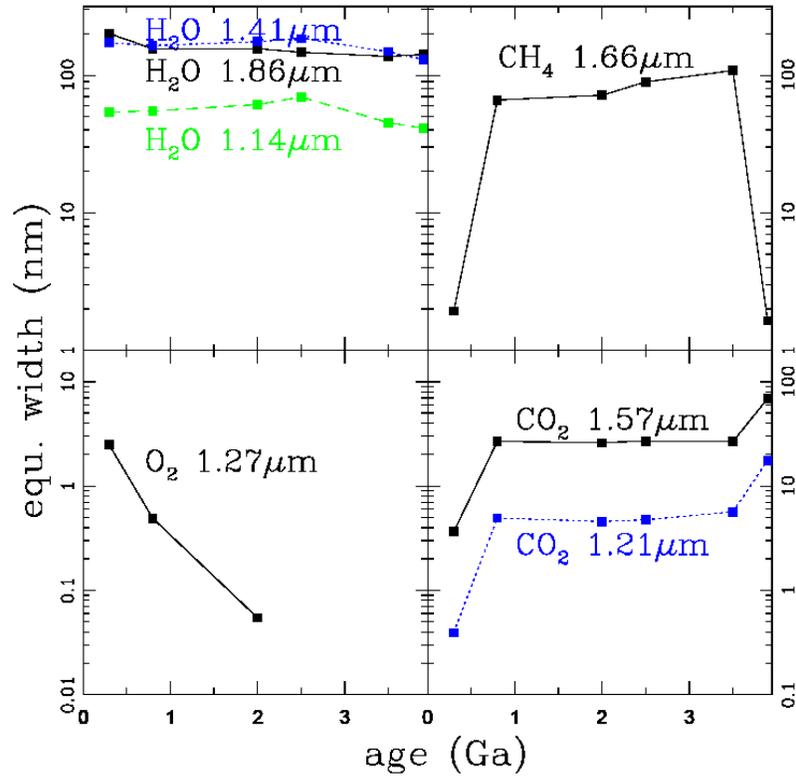

**Fig. 18:** Equivalent width (nm) of the main spectral features of atmospheric compounds over geological times in the near-infrared with no clouds.

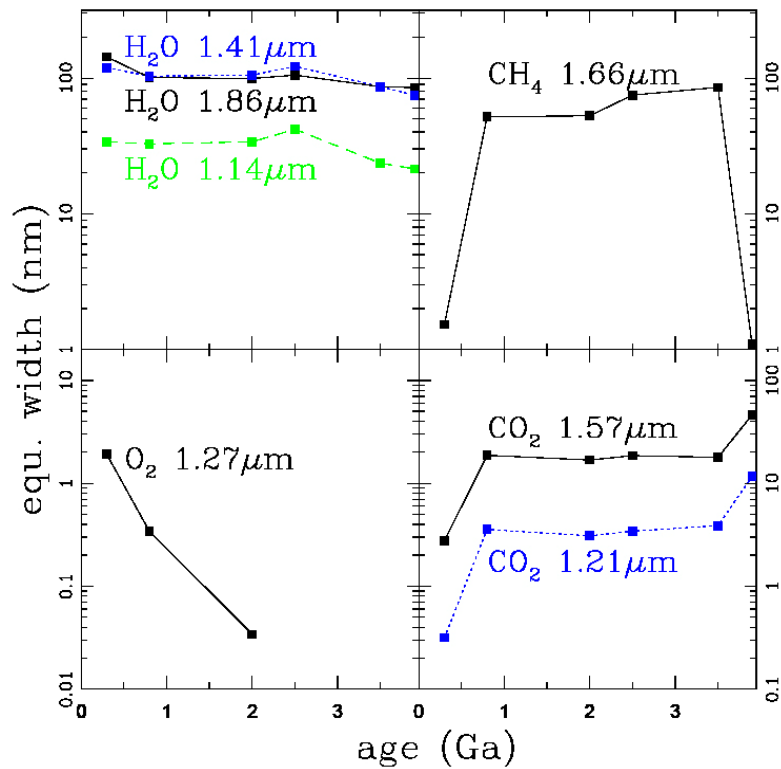

**Fig. 19:** Equivalent width (nm) of the main spectral features of atmospheric compounds over geological times in the near-infrared with Earth clouds.

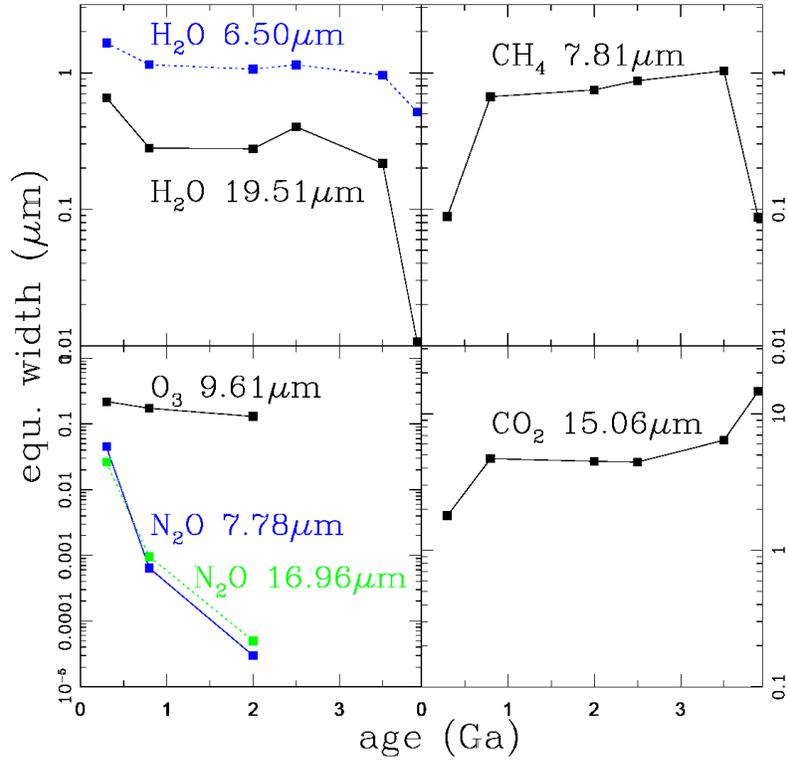

**Fig. 20: Equivalent width (μm) of the main spectral features of atmospheric compounds over geological times in the thermal infrared with no clouds.**

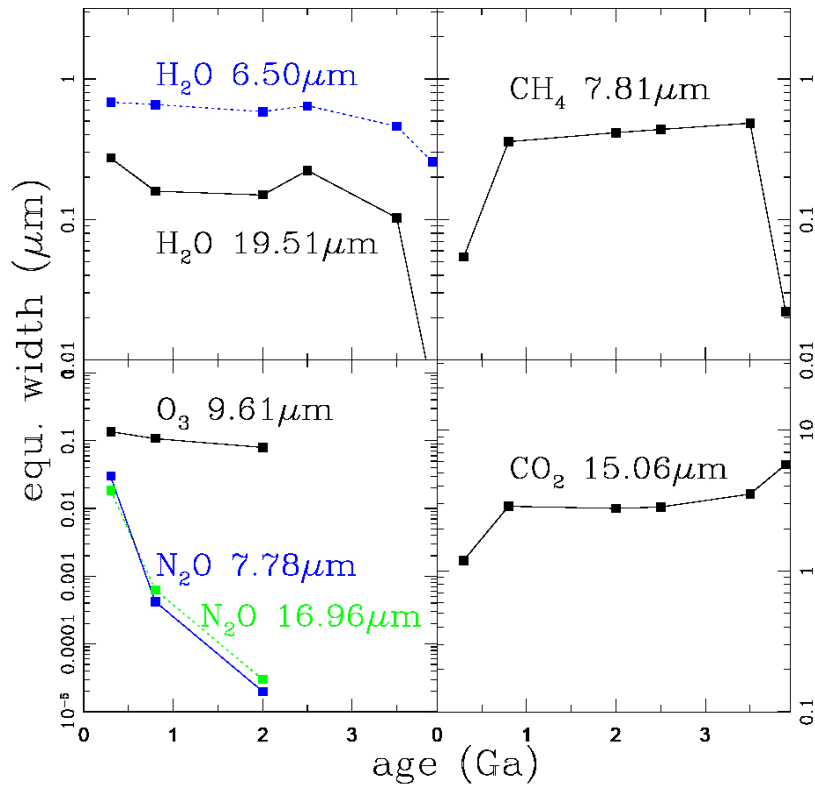

**Fig.21: Equivalent width (μm) of the main spectral features of atmospheric compounds over geological times in the thermal infrared with Earth clouds.**

## S.2 RESOLUTION

Here we show the spectral resolution ($\lambda/\Delta\lambda$) needed for optimal detection of each spectral feature, over geological time, and in the visible (see Fig. 22), near-infrared (see Fig. 23), and thermal infrared (see Fig. 24), as given in Tables 8 through 10. To detect a spectral feature with optimum signal to noise requires that the FWHM of the spectrometer should be approximately equal to the FWHM of the spectral features. Here $\lambda$ is the central wavelength of a feature, and $\Delta\lambda$ is the FWHM of the feature after it has been smeared sufficiently to blend any sharp lines yet still retain its essential overall shape.

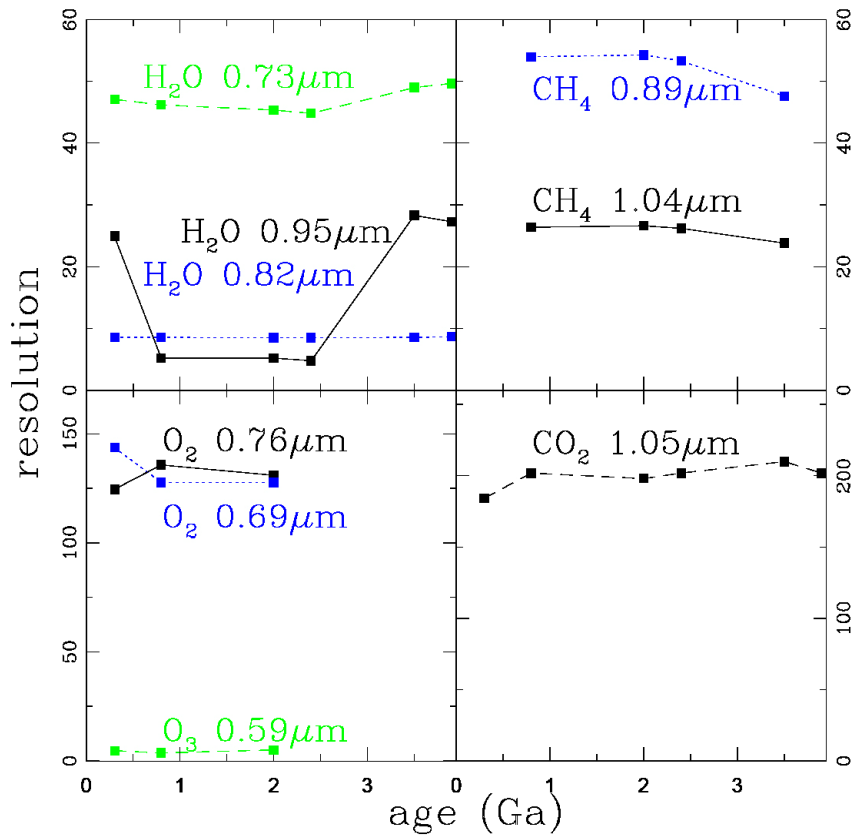

**Fig. 22: Resolution ($\lambda/\Delta\lambda$) needed to match the main spectral features of atmospheric compounds over geological time in the visible with Earth clouds.**

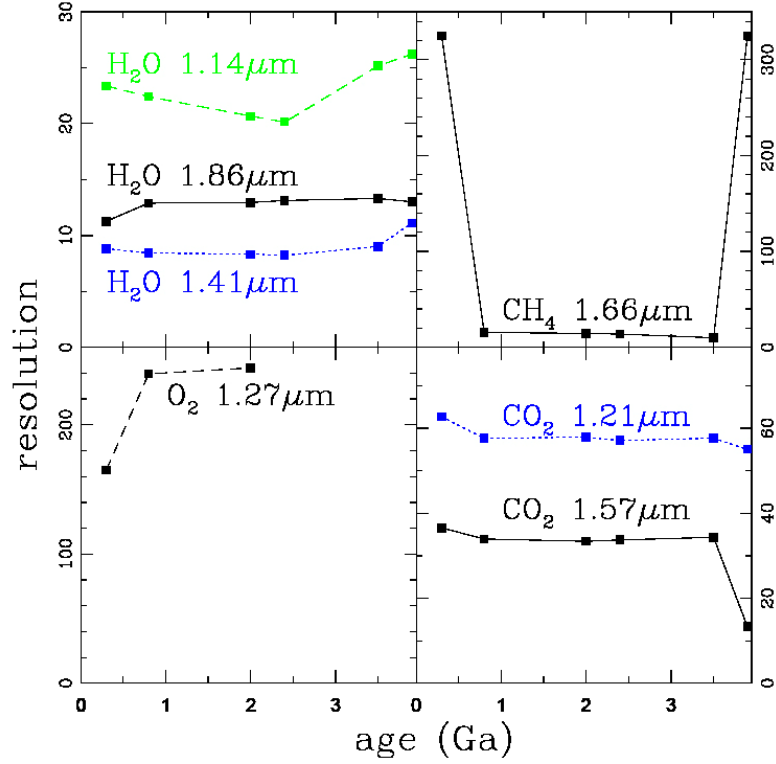

**Fig. 23: Resolution ($\lambda/\Delta\lambda$) needed to match the main spectral features of atmospheric compounds over geological time in the near-infrared with Earth clouds.**

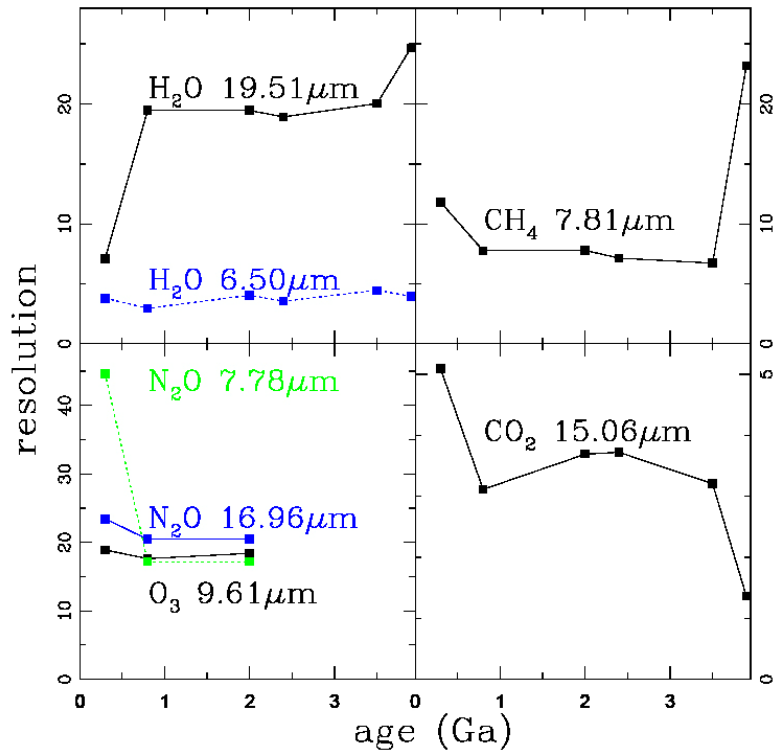

**Fig. 24: Resolution ($\lambda/\Delta\lambda$) needed to match the main spectral features of atmospheric compounds over geological times in the thermal infrared with Earth clouds.**